\documentclass[aps,showpacs,epsf,twocolumn]{revtex4}
\usepackage{graphicx}
\usepackage{dcolumn}
\usepackage{amsmath}
\usepackage{amssymb}

\usepackage[colorlinks]{hyperref}
\hypersetup{citecolor=blue}

\usepackage{bm}
\usepackage{epstopdf}
\usepackage{amsthm}
\usepackage{float}

\def\ie{i.e., }

\def\epsilon{\varepsilon}

\def\i{{\mbox{i}}}

\def\Im{{\mbox{Im}}}
\def\Re{{\mbox{Re}}}

\def\epsilon{\varepsilon}

\def\du{{\mbox d}u~}
\def\xkpd{x_k^{\text{PD}}}
\def\xkc{x_k^{\text{C}}}

\def\p(u){\frac{e^{-u/b}}{b}}

\newcommand*{\Comb}[2]{{}^{#1}C_{#2}}

\begin{document}

\title{How clock heterogeneity affects synchronization and can enhance stability}

\author{Nirmal Punetha$^1$ and Lucas Wetzel$^{1,2}$} 
\affiliation{$^1$Max Planck Institute for the Physics of Complex Systems, N\"othnitzer Stra{\ss}e 38, 01187 Dresden, Germany}
\affiliation{$^2$Center for Advancing Electronics Dresden, cfaed, W\"{u}rzburger Stra{\ss}e 46, 01187, Germany}

\begin{abstract}
The production process of integrated electronic circuitry inherently leads to large heterogeneities on the component level. 
For electronic clock networks this implies detuned intrinsic frequencies and differences in coupling strength and the characteristic time-delays associated with signal transmission, processing and feedback.
Using a phase-model description, we study the effects of such component heterogeneity on the dynamical properties of synchronization in networks of mutually delay-coupled Kuramoto oscillators.
We test the theory against experimental results and circuit-level simulations in a prototype system of mutually delay-coupled electronic clocks, so called phase-locked loops. 
Contrary to the hindering effects of component heterogeneity for the synchronization in hierarchical networks, we show that clock heterogeneities can enhance self-organized synchronization in networks with flat hierarchy.
That means that beyond the optimizations that can be achieved by tuning homogeneous coupling strengths, time-delays and loop-filter cut-off frequencies, heterogeneities in these system parameters enable much better optimization of perturbation decay rates, the stabilization of synchronous states and the tuning of phase-differences between the clocks.
%
Our theory enables the design of custom-fit synchronization layers according to the specific requirements and properties of electronic systems, such as operational frequencies, phase-relations and e.g. transmission-delays.
These results are not restricted to electronic systems, as signal transmission, processing and feedback delays are common to networks of spatially distributed and coupled autonomous oscillators.   
\end{abstract}
\maketitle

\section{Introduction} \label{sec:intro}

In modern electronics the coordination of complex systems and processes in time is necessary for a defined system behavior, efficient operation and reliable parallel information processing \cite{Corbett2012,Brewer2017,Shrit2017}. 
To establish such coordination via a global time reference for spatially distributed clock elements can be a challenging task, especially at high frequencies and in the presence of signal transmission- and processing-delays.
Such delays are induced by finite signal propagation speeds in transmission lines and due to signal processing, e.g., filtering.
Furthermore, integrated electronic systems display a considerable degree of heterogeneity in their component characteristics due to, e.g., the production process of semiconductor technology~\cite{Verma2009,Alioto2010,Onabajo2012}.
This has far reaching consequences for the architectures of electronic systems, requiring their functionality to be robust against such heterogeneity.
Example systems are global and indoor positioning, large antenna, radar and sensory arrays, multi-processor computer architectures, terahertz based technology and databases on the internet~\cite{Corbett2012}.
A common approach to synchronization in such systems is to entrain imprecise electronic clocks, so called phase-locked loops (PLLs)~\cite{Yuldashev2015}, hierarchically with a dedicated and precise reference clock (usually a quartz).
Such references feed their signal unidirectionally into a clock-tree that becomes increasingly complicated as the system size grows~\cite{Ho2001,Mensink2010}.
Therefore such systems are often synchronized only locally by \textbf{g}lobally \textbf{a}synchronous, \textbf{l}ocally \textbf{s}ynchronous operations (\textit{GALS})~\cite{Yu2009}.
A novel approach to the synchronization of large spatially distributed electronic systems is to allow the formation of self-organized synchronous states \cite{Orsatti2008,Javidan2011,Pollakis2014,Koskin2018}.
This is inspired by robust self-organized synchronization without hierarchical structures as found in biological systems, where synchronization is achieved robustly in highly noisy environments with strong heterogeneities and in the presence of considerable time-delays \cite{Morelli2009,Oates2012}.
So far, such networks have been studied analytically for homogeneous clock networks and and have been tested experimentally with weakly heterogeneous electronic prototype clocks \cite{Pollakis2014,Jorg2015,Wetzel2017}.
To enable this technology for applications in electrical engineering, it is important to understand the consequences of heterogeneity on the collective self-organized dynamics in finite-size systems \cite{Koskin2018}.
Furthermore, studying synchronization of heterogeneous clocks is relevant beyond electronic systems and can provide insight in, e.g. power grid and biological neural networks, consisting of strongly heterogeneous units \cite{Balasubramanian2015,Tenti2016,Kehl2017}.
In this paper we study and analyze the effects of heterogeneity on the synchronization dynamics of mutually delay-coupled electronic clocks.
We use a Kuramoto-type model, i.e., networks of delay-coupled phase oscillators with node dynamics that include processing and inversion of signals and delayed feedback \cite{Kuramoto1984,Strogatz2000}.

The paper is organized as follows.
We introduce a model for heterogeneous networks of non-identical mutually delay-coupled digital electronic clocks with signal filtering in Sec.~\ref{sec:2hetoPLL}. 
For such systems we calculate the frequencies and phase configurations of synchronized states, and analyze their stability as a function of different heterogeneous system parameters in Sec.~\ref{sec:sync-sol-new}.
In Sec.~\ref{sec:results} we discuss the effects of these heterogeneities and compare to the case of networks of identical clocks.
We present how the different types of time-delays and their interplay affects the dynamical properties of synchronization in such systems.
We summarize in Sec.~\ref{sec:discuss}, connect the results to modern electronic applications and components, and discuss the potential of self-organized synchronization for microelectronic systems and processes.

\section{Mutually delay-coupled electronic clocks} \label{sec:2hetoPLL}

\begin{figure}[t]
\includegraphics[scale=.305, angle=0]{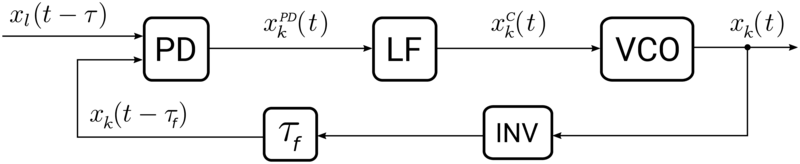}
 \caption{Simplified schematic of a phase-locked loop unit. \emph{PD} denotes the phase detector, \emph{LF} the loop filter, \emph{VCO} the voltage controlled oscillator, 
 \emph{INV} the inverter in the feedback loop and 
 $\tau_f$ the feedback-delay element.}
 \label{fig:sketchPLL}
\end{figure}

We consider a system of mutually delay-coupled electronic clocks, so called phase-locked loops (PLLs) \cite{Meyr1986,Gardner2005}.
Each PLL of the clock networks considered here consists of a phase-detector (PD), a loop-filter (LF), a voltage controlled oscillator (VCO), 
an inverter (INV) and a feedback-delay component in the feedback path, see Fig.~\ref{fig:sketchPLL}.
Heterogeneity in these components manifests in heterogeneous intrinsic frequencies, filter cut-off frequencies, transmission- and feedback-delays.
The PLLs are mutually connected, i.e., each VCO sends its output signal $x_k(t)$ to at least one other and receives time-delayed signals $x_l(t-\tau_{kl})$ from at least another oscillator in the network.
The PD generates a signal that contains information about the phase relations between the input signals $x_l(t-\tau_{kl})$ from other PLLs and the internal feedback signals $x_k(t-\tau_{kl}^f)$ 
%
\begin{equation} \label{eq:pd-output}
\xkpd(t) = \frac{1}{n_k}\sum_{l=1}^{N}\,c_{kl}\,h^{\rm PD}\left[ \phi_l(t-\tau_{kl}), \phi_k(t-\tau_{kl}^f) \right],
\end{equation}
where $h^{\rm PD}(\,\cdot\,)$ denotes $2\pi$-periodic coupling functions that depend on the type of phase detection at the PD (e.g. XOR or analog multiplier),
$\phi_l(t-\tau_{kl})$ denotes the phases of the input signals from other PLLs, delayed by transmission-delays $\tau_{kl}$, 
$\phi_k(t-\tau_{kl}^f)$ denotes the phase of the feedback signal, delayed by a feedback-delay $\tau_{kl}^f$, 
and $n_k=\sum_l \,c_{kl}$ is the node degree and denotes the number of the nodes' input signals.
The $c_{kl}$, equal to one if there is a connection between PLL $l$ and PLL $k$ and zero otherwise, denote the elements of the adjacency matrix that specifies the coupling topology.
The LF then processes the PD signal which yields the filtered control signal $\xkc(t)$
\begin{equation} \label{eq:lf-output}
\xkc (t) = \int_0^{\infty}\du\,p_{k}(u)\,\xkpd(t-u),
\end{equation}
\noindent
where $p_{k}(u)$ denotes the impulse response of the LF. 
Usually the LF is implemented as a RC low-pass.
Its transfer-function in Laplace-domain is represented in time-domain by the Gamma-distribution~\cite{Mancini2003}.
A large class of LFs can hence be modeled as $p_{k}(u)=p(u;a_k,b_k)$ \cite{Mancini2003}, where 
\begin{equation} \label{eq:pu01}
\begin{split}
p(u;a,b) = u^{(a-1)} \frac{e^{-u/b}}{b^{a} ~\Gamma(a) },
~\rm{and}~\int_0^{\infty} \du p(u;a,b) = 1.
\end{split}
\end{equation}
The LF is characterized by the order $a_{k}$ of the filter, and the scale parameter $b_{k}$ which are related to the cut-off frequency of the LF as $\omega_{k}^c = 2 \pi f_{k}^c = 1/(a_{k} b_{k})$. 
The VCO is operated such that it responds linearly to the control signal $\xkc(t)$
\begin{equation} \label{eq:vco-output}
\dot{\phi}_k = \omega_k^0 + K_{k}^{\rm VCO} \xkc (t),
\end{equation}
where $k=1,\dots,N$ indexes the PLLs in the network, $\omega_k^0$ denotes the intrinsic frequency and $K_{k}^{\rm VCO}$ denotes the input sensitivity of VCO $k$. 
Using Eq.~(\ref{eq:pd-output}), Eq.~(\ref{eq:lf-output}) and Eq.~(\ref{eq:vco-output}) we obtain the phase model
\begin{equation} \label{eq:sys-sine}
\dot{\phi}_k(t) = \omega_k + \frac{K_{k}}{n_k} \sum_{l=1}^N c_{kl} \int_0^{\infty} \du p_{k}(u) \cdot X_{kl}(t,u,\tau_{kl},\tau_{kl}^f),
\end{equation} 
where we defined the coupling strength $K_k=K_k^{\rm VCO}/2$, and $X_{kl} = h\left[ \phi_l(t-u-\tau_{kl})-\phi_k(t-u-\tau_{kl}^f) \right]$ denotes the low frequency components of $h^{\rm PD}[\,\cdot\,]$. 
We here consider the high frequency components to be ideally filtered by the LF \cite{Pollakis2014}. 
In the case of digital signals and an XOR phase detector, as will be shown in the section with the experimental results, the coupling function $h[\,\cdot\,]$ is a triangular function ~\cite{Wetzel2017}.
In the following, we consider a cosine coupling function associated with analog signals and a multiplier PD, see~\cite{Pollakis2014}.
Note that for an LF impulse response peaked at zero, $p(u)=\delta(u)$, a sinusoidal coupling function $h(\,\cdot\,)$, and $\tau_{kl}=\tau_{kl}^f=0$, Eq.~(\ref{eq:sys-sine}) reduces to a Kuramoto model of coupled phase oscillators with heterogeneous intrinsic frequencies \cite{Schuster1989,Acebron2005}.

\section{Synchronized solutions and linear stability} \label{sec:sync-sol-new}

We now consider a system of two delay-coupled phase-locked loops.
This minimal system is suitable to exemplify the implications of component heterogeneities for the dynamics of self-organized synchronization.
The general case of $N$ delay-coupled heterogeneous PLLs is shown in the Supplementary material~\ref{supp-mat-Nosc}.
The instantaneous frequencies of the two analog delay-coupled PLLs are given by  
\begin{equation}\label{eq:n=2-system}
\begin{split}	
	\dot{\phi}_{1,2}(t) &= \omega_{1,2} + K_{1,2} \int_0^{\infty} \du p_{1,2}(u) \\
	&\cos\left[ \phi_l(t-u-\tau_{12,21})-\phi_k(t-u-\tau_{12,21}^f) \right].
\end{split}
\end{equation}
We are interested in phase-locked synchronized states, i.e., all oscillators evolve with same collective frequency $\Omega$ and have constant phase-lag $\beta$ between them 
\begin{equation}\label{eq:n=2-sync-ansatz}
\begin{split}	
	\phi_1 = \Omega t;~~\phi_2 = \Omega t + \beta.
\end{split}
\end{equation}
%
%
In this case synchronized solutions with global frequency $\Omega$ and phase difference $\beta$ are given by the following transcendental equations~(derivation see Appendix \ref{sec:app-sync-sol-N2})
\begin{equation}\label{eq:n=2-all-hetero-sync-sol}
\begin{split}
\Omega &= \bar{\omega} + \bar{K} \cos(\Omega \bar{\tau}_e) \cos(B) - \frac{\Delta K}{2} \sin(\Omega \bar{\tau}_e) \sin(B),\\
 \beta &= - \frac{\Omega \Delta \tau_e}{2} + \sin^{-1}\left( \frac{\Delta\omega}{H_1}\right) + 
\sin^{-1}\left( \frac{\Delta{K} \cos(\Omega \bar{\tau}_e)}{H_1}\right), 
\end{split}
\end{equation}
where $B = (\Omega \Delta{\tau_e})/2 + \beta$ and
\begin{equation}
H_1 = \sqrt{ \left(2 \bar{K} \sin(\Omega \bar{\tau}_e) \right)^2 + \left( \Delta{K} \cos(\Omega \bar{\tau}_e) \right)^2}.
\end{equation} 
The heterogeneous parameters, $\omega_{1,2}$, $\tau_{12,21}$, $\tau^f_{12,21}$, $K_{1,2}$, $\omega_{c1,c2}$ are written in terms of their mean $\bar{x} = (x_1+x_2)/2$ and difference $\Delta x = x_2-x_1$. 
Parameter symbols without a subscript indicate identical parameters for both oscillators.
%
We also defined $\bar{\tau}_e = (\bar{\tau} - \bar{\tau}^f)$, and $\Delta{\tau}_e = (\Delta{\tau} - \Delta{\tau}^f)$. 
For the case of identical oscillators, in- and antiphase synchronized states exist, characterized by a common global frequency $\Omega$ and a constant phase difference $\beta$ which is either zero or $\pi$, respectively.
How heterogeneous parameters affect the existence of synchronized solutions, the global frequencies and phase configurations given by Eq.~\eqref{eq:n=2-all-hetero-sync-sol} will be discussed in detail in Sec.~\ref{sec:results}. 

The stability of such solutions can be determined by analyzing the response to small perturbations.
More precisely, we can look at the perturbation mode related to the phase-difference between the oscillations of two PLL clocks.
The characteristic equation governing the exponential growth or decay rate $\lambda$ of these perturbations is given by (derivation see appendix~\ref{sec:app-stab})
\begin{equation}\label{eq:n=2-all-hetero-stab-con}
\begin{split}
\frac{\lambda^2}{\alpha_{12} \hat{p}_{1}(\lambda) \alpha_{21} \hat{p}_{2}(\lambda) e^{-2\lambda \bar{\tau}^{f}}}\hspace{4cm}\\
+ \lambda 
\left( 
\frac{1}{ \alpha_{12} \hat{p}_{1}(\lambda) e^{-\lambda \tau^{f}_{1}} } +
\frac{1}{ \alpha_{21} \hat{p}_{2}(\lambda) e^{-\lambda \tau^{f}_{2}} } 
\right)\\  
- \left( e^{-2\lambda (\bar{\tau} - \bar{\tau}^f)} - 1 \right)
=0,
\end{split}
\end{equation}
\noindent
where the $\lambda$ denote the eigenvalues of the perturbation modes, $\hat{p}(\lambda) = (1+\lambda b_{k})^{-1}$ denotes the Laplace transform of the impulse response function introduced in Eq.~(\ref{eq:pu01}), 
and $\alpha_{12} = K_1 ~h^{\prime} [ ( -\Omega(\tau_{12}-\tau^{f}_{1}) + \beta ) ]$, $\alpha_{21} = K_2 ~h^{\prime} [ ( -\Omega(\tau_{21}-\tau^{f}_{2}) - \beta ) ]$ 
with $h^{\prime}$ being the derivative of the coupling function with respect to its argument. 
The eigenvalue ($\lambda_{\rm max} = \sigma + i \gamma$) with the largest real part $\sigma$ dominates the long-term behavior and determines the stability of synchronized states. 
For $\sigma > 0$ perturbations to synchronized states grow and the solution is unstable, whereas $\sigma < 0$ implies linear stability and that perturbations decay at a characteristic time-scale $t_c=-\sigma^{-1}$. 
The case of $\sigma = 0$ denotes marginal stability, \ie perturbations neither grow or decay. 
This is always a solution to Eq.~\eqref{eq:n=2-all-hetero-stab-con} and represents an equal shift of the phase of each oscillator.
The imaginary part $\gamma  = \Im(\lambda_{\rm max})$ denotes the frequency of the perturbation response.
Since the stability depends on the frequency and phase-configurations, the delays and cut-off frequencies, it can also be modified when heterogeneities are introduced.

\section{Effects of heterogeneities on synchronization} \label{sec:results}
In this section, we systematically analyze the effects of heterogeneity in the different clock components. 
The results are compared to those obtained from an experimental setup of two identical delay-coupled PLLs. 
We show how the notion of in- and antiphase synchronized states, observed for identical oscillators, becomes blurred when heterogeneities in the intrinsic frequencies, coupling strength, transmission and feedback-delays are introduced. 
For those cases we define so called \textit{asymptotic-inphase} and \textit{-antiphase} synchronized states that approach the in- and antiphase synchronized states for identical clocks as the heterogeneities approach zero. 
In the following sections we will introduce the heterogeneities gradually, starting with heterogeneity in the intrinsic frequencies, followed by the heterogeneities in the other parameters.

\subsection{Heterogeneous intrinsic frequencies} \label{subsec:detuning}
Here we analyze how synchronization in a system of two delay-coupled PLLs is affected by heterogeneous intrinsic frequencies. 
The results that we obtain for the second-order Kuramoto-model are identical to the ones obtained for the first-order Kuramoto-model with time-delayed coupling and no signal filtering \cite{Schuster1989}.
We assume that signal transmission-delays, coupling strengths and impulse response functions of the LFs for the two oscillators are equal, \ie $\tau_{12} = \tau_{21} = \tau$, $K_1=K_2=K$, $p_1(u) = p_2(u) = p(u)$. 
The feedback-delays are set to zero, $\tau^{f}_{12} = \tau^{f}_{21} = 0$.
Under these assumptions we find for the frequencies in Eq.~\eqref{eq:n=2-all-hetero-sync-sol}
\begin{equation}\label{eq:n=2-sfreq-sol-h1}
\begin{split}
\Omega = \bar{\omega} \pm K \cos \left( -\Omega {\tau} \right) 
\sqrt{ 1 - \left( \frac{\Delta\omega}{2 K \sin \left( \Omega {\tau} \right)} \right)^2 }, 
\end{split}
\end{equation}
\noindent
and the corresponding phase differences $\beta$
\begin{equation}\label{eq:n=2-sphase-sol-h1}
\beta = 
\begin{cases}
\sin^{-1} \left( \frac{\Delta\omega}{2 K \sin \left( \Omega {\tau} \right)} \right),
~~~~~~~\mbox{if}~~\sin \left( \Omega {\tau} \right)>0, \\
\\
\pi - \sin^{-1} \left( \frac{\Delta\omega}{2 K \sin \left( \Omega {\tau}) \right)} \right),
~~\mbox{if}~~\sin \left( \Omega {\tau}  \right)<0. \\
\end{cases}
\end{equation}
\noindent
This quantifies the phase differences $\beta$ as a function of the detuning $\Delta \omega$ between the intrinsic frequencies.
The phase differences are no longer $0$ (inphase) or $\pi$ (antiphase) as in the case of identical oscillators.
Instead, new phase configurations emerge due to the detuning. 
They depend on the ratio of frequency differences to coupling strength $\propto (\Delta\omega/K)$, see Eq.~\eqref{eq:n=2-sphase-sol-h1}. 
\begin{figure}
  \includegraphics [scale=0.305, angle=0]{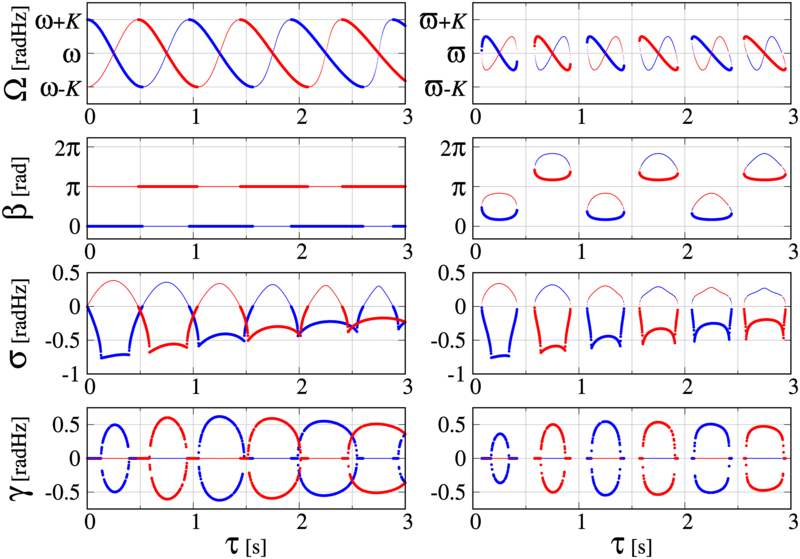}
  \caption{(Color online) 
    Global frequency $\Omega$, the phase-difference $\beta$, the perturbation response rate $\sigma$ and the corresponding modulation frequency $\gamma$ as a function of the transmission-delay $\tau$ for two mutually delay-coupled PLLs with $K=0.25\,\rm{radHz}$ and $\omega_c=0.25\times\bar{\omega}\,\rm{radHz}$.
    The left column shows the results with identical ($\omega_{1,2} = 2\pi\,\rm{~rad Hz}$) and the right column the results for heterogeneous intrinsic frequencies ($\omega_{1,2} = (1\mp0.02)2\pi\,\rm{radHz}$).
    The blue (dark gray) and red (light gray) curves correspond to the inphase and antiphase ($\Delta\omega = 0$) or asymptotic-inphase and -antiphase 
    ($\Delta\omega = 0.04\times 2 \pi\,\rm{radHz}$) synchronized states.
    The thick curves denote stable solutions (from Eq.~\eqref{eq:n=2-stabcon2-h1}, with $\sigma=\Re(\lambda_{max})<0$) and the thin curves denote unstable solutions.}
  \label{fig:new-hf-fig11}
\end{figure}
\begin{figure}[!th]
  \includegraphics [scale=0.305, angle=0]{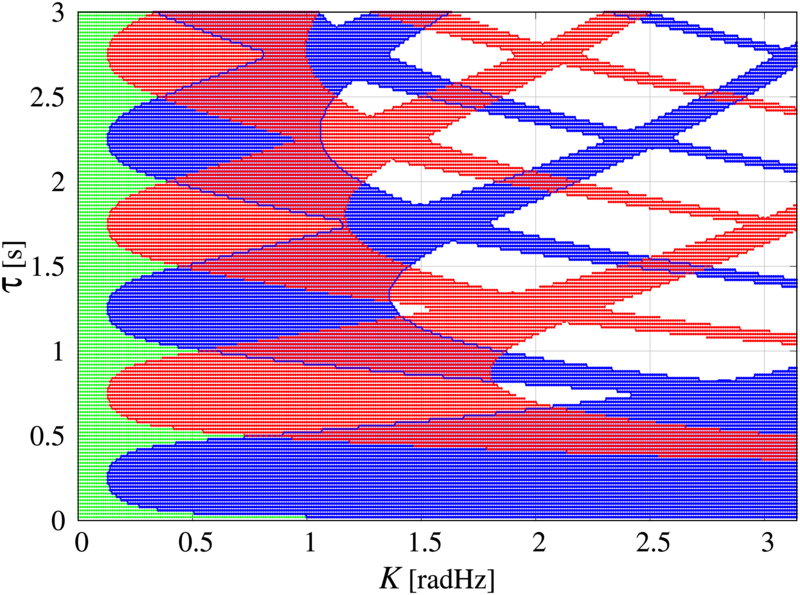}
  \caption{(Color online)
  Existence and stability of synchronized solutions for two delay-coupled PLLs in the parameter plane of coupling strength and transmission-delay with detuned intrinsic frequencies $\omega_{1,2} = (1\mp0.02)\times 2\pi \rm{~rad Hz}$, $\omega_c = 0.25\times \bar{\omega} \rm{~rad Hz}$.
  Asymptotic-inphase and -antiphase synchronized solutions are denoted by the blue and red colored regions, respectively. 
  In overlapping regions both solutions are stable. 
  The green region represents the parameter values where synchronized states do not exist due to the detuning $\Delta \omega$. 
  The instabilities induced by the filtering process lead to the blank (white) regions where synchronized states do exist but are not stable.
  In those regimes new solutions with modulated frequencies emerge.}
  \label{fig:new-hf-fig14}
\end{figure}
The detuning also implies, see Eq.~\eqref{eq:n=2-sfreq-sol-h1}, that synchronized solutions exist only if 
\begin{equation}\label{eq:n=2-sfreq-sol-restriction}
\begin{split}
\sin^2 \left( \Omega {\tau} \right) \geq \left( \frac{\Delta\omega}{2 K} \right)^2.
\end{split}
\end{equation}
\noindent
Therefore, synchronized solutions do not exist below a critical value of the coupling strength $K_{\rm c} = \Delta\omega/2$ for any delay value, that is, if $\vert \Delta \omega/2K \vert > 1$. 

From Eq.~\eqref{eq:n=2-all-hetero-stab-con} the linear stability of such synchronized solutions is given by
\begin{equation}\label{eq:n=2-stabcon2-h1}
  \begin{split}
  \frac{\lambda^2}{\hat{p}^{2}(\lambda)} + 
  \left( \alpha_{21} + \alpha_{12} \right) \frac{\lambda}{\hat{p}(\lambda)} 
  - \alpha_{12} \alpha_{21} \left( e^{-2\lambda {\tau}} - 1 \right)
  =0,
  \end{split}
  \end{equation}
where $\alpha_{12,21} = -K\sin(-\Omega\tau \pm \beta)$ and $\hat{p}(\lambda) = (1+\lambda b)^{-1}$. 
In Fig.~\ref{fig:new-hf-fig11} synchronized solutions in a system without heterogeneity (left column) and with heterogeneity (right column) in the intrinsic frequencies are shown. 
The coloring identifies asymptotic-inphase and -antiphase states, blue and red, respectively.
In the right column of Fig.~\ref{fig:new-hf-fig11} for heterogeneous frequencies, we observe regimes for which no synchronized solutions exist, since Eq.~\eqref{eq:n=2-sfreq-sol-restriction} is not satisfied.
These gaps (windows) appear due to the detuning and are absent when $\Delta\omega = 0$, see left column. 
In that case, we find that pairs of synchronized solutions go through a saddle-node bifurcation as the magnitude of the detuning $\Delta\omega$ is increased from $\Delta\omega = 0$, see Supplementary material Fig.~\ref{fig:new-hf-fig13}.
Furthermore, for identical oscillators, non-generic solutions with coinciding frequencies exist for specific parameter values $\Omega\tau_e = n\pi, \;n \in \mathbb{N}_0^+$, see Appendix~\ref{sec:app-sync-sol-N2}. 
These solutions split up due to the symmetry breaking, i.e., additional solutions emerge as the intrinsic frequencies are detuned. 
The range of possible global frequencies decreases.
For identical oscillators, the global frequencies lie in the range $\Omega \in [\omega-K, \omega+K]$, whereas for detuned frequencies this range decreases by $\Delta\omega$, \ie $\Omega \in [\bar{\omega}-K+\Delta\omega/2, \bar{\omega}+K-\Delta\omega/2]$. 
That is, the system can only self-organize to synchronized states with global frequencies that can be reached by all oscillators in the network, see experimental results in \cite{Wetzel2017}. 

Linear stability of such synchronized states is also affected by the detuning as it depends on the values $\Omega$ and $\beta$ of the synchronized solutions.
In Fig.~\ref{fig:new-hf-fig14} we provide an overview of existence and linear stability of synchronized solutions in parameter space, \ie the ($K, \tau$)-plane. 

\subsection{Heterogeneity in transmission-delays} \label{subsec:transmission-delay}

\begin{figure}[t!]
\includegraphics [scale=0.305, angle=0]{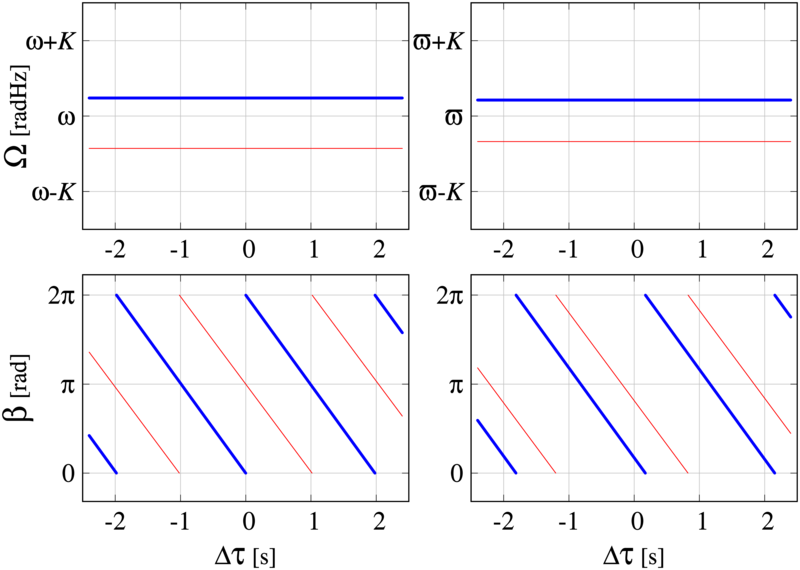}
 \caption{(Color online) 
Global frequency $\Omega$ and phase-difference $\beta$ as a function of the delay-difference $\Delta\tau$ for $N=2$ delay-coupled PLLs with identical $\omega_{1,2} = 1 \times 2\pi \rm{~rad Hz}$ (left column) and heterogeneous intrinsic frequencies $\omega_{1,2} = (1\mp0.02)\times 2\pi \rm{~rad Hz}$ (right column). 
Mean transmission-delay $\bar{\tau}$ is fixed at $1.2 \rm{~s}$, $K=0.25 \rm{~rad Hz}$ and $\omega_c = 0.25\times \bar{\omega} \rm{~rad Hz}$. 
The blue (dark gray) and red (light gray) curves correspond to the inphase and antiphase ($\Delta\omega=0$) or asymptotic-inphase and -antiphase ($\Delta\omega \neq 0$) synchronized states. 
The thick and thin lines denote stable and unstable solutions, respectively. 
}
\label{fig:new-hf-dd-fig1}
\end{figure}

Now we add heterogeneous transmission-delays, \ie $\tau_{12} \neq \tau_{21}$ to the system with detuned intrinsic frequencies. 
The frequencies and phase configurations then are
%
\begin{equation} \label{eq:n=2-sfreq-sol-h2}
\begin{split}
\Omega = \bar{\omega} \pm K \cos \left( -\Omega \bar{\tau} \right) 
\sqrt{1 - 
\left( \frac{\Delta\omega}{2 K \sin \left( \Omega \bar{\tau} \right)} \right)^2 }.
\end{split}
\end{equation}
and 
\begin{equation}\label{eq:n=2-sphase-sol-h2}
\beta = 
\begin{cases}
&-\frac{\Omega \Delta\tau}{2} + 
\sin^{-1} \left( \frac{\Delta\omega}{2 K \sin \left( \Omega \bar{\tau} \right)} \right) \vspace{0.25cm} \\
&~~~~~~~~~~~~~~\mbox{if}~~\sin \left( \Omega \bar{\tau} \right)>0, \\
\\
&-\frac{\Omega \Delta\tau}{2} +
\pi - \sin^{-1} \left( \frac{\Delta\omega}{2 K \sin \left( \Omega \bar{\tau} \right)} \right)\vspace{0.25cm} \\ 
&~~~~~~~~~~~~~~\mbox{if}~~\sin \left( \Omega \bar{\tau} \right)<0, \\
\end{cases}
\end{equation}
see Eq.~\eqref{eq:n=2-all-hetero-sync-sol}. 
As in the last section we set feedback-delays to zero, and study the effects of heterogeneous transmission-delays. 
The stability is given by the characteristic Eq.~\eqref{eq:n=2-all-hetero-stab-con} 
\begin{equation}\label{eq:n=2-stabcon2-h2}
  \begin{split}
  \frac{\lambda^2}{\hat{p}^{2}(\lambda)} + 
  \frac{\lambda \left( \alpha_{21} + \alpha_{12} \right)}{\hat{p}(\lambda)} 
  - \alpha_{12} \alpha_{21} \left( e^{-2\lambda \bar{\tau}} - 1 \right)
  =0,
  \end{split}
  \end{equation}
where $\alpha_{12} = -K\sin(-\Omega \tau_{12} + \beta)$,  $\alpha_{21} = -K\sin(-\Omega \tau_{21} - \beta)$ and $\hat{p}(\lambda) = (1+\lambda b)^{-1}$. 
These equations show that the frequencies of synchronized states depend only on the mean delay value $\bar{\tau} = (\tau_{12}+\tau_{21})/2$ but not the delay difference $\Delta\tau=\tau_{21}-\tau_{12}$. 
The same is true for the characteristic equation which only depends on the mean delay $\bar{\tau}$, see~\cite{eqv-delay}. 
Therefore the global frequencies of synchronized states and their stability remains the same as long as the mean delay value $\bar{\tau}$ is unchanged, see Fig.~\ref{fig:new-hf-dd-fig1}. 
However, the effect of heterogeneous delays becomes significant for the phase-differences $\beta$ with its linear dependence on $\Delta \tau$. 
As a result, the phase difference changes monotonically with $\Delta\tau$ without affecting the frequency and stability of synchronized states. 
This is an interesting results since it allows the phase difference to be tuned to arbitrary values at a fixed frequency. 
In Sec.~\ref{sec:exp-sim} we show this in experimental results for delay-coupled digital PLLs.

\subsection{Heterogeneity in signal filtering parameters} \label{subsec:filtering}

Analyzing Eq.~\eqref{eq:n=2-all-hetero-sync-sol}, we find that the filtering process has no effect on the frequencies and phase differences of synchronized solutions.  
However, the stability of synchronized solutions depends on the cut-off frequencies $\omega_{1,2}^c$ and the order $a_{1,2}$ of the LFs via the expression 
$\hat{p}_{i}(\lambda) = (1 + \lambda/(a_{i}\omega_{i}^c))^{-a_{i}}$ for $i=\{1,2\}$, see Eq.~\eqref{eq:n=2-all-hetero-stab-con}. 
We consider filters of first order ($a_i=1$) and set $\omega_{1,2}^c = \bar{\omega}^c \mp \Delta\omega^{c}/2$. 
Note that for zeroth order filters the phase equations reduce to a first order Kuramoto model with delayed coupling, see Supplementary material \ref{supp-mat-second-order-km}. 
For delay-coupled identical Kuramoto oscillators, Yeung, Earl and Strogatz~\cite{Yeung1999, Earl2003} provide a necessary and sufficient condition for the stability of inphase synchronized solutions, given by $K h^{\prime}[-\Omega \tau] > 0$.
However, in coupled PLL systems, first or higher order LFs have significant effects on stability.
In these cases, the stability criterion for systems without a LF is still sufficient but not necessary ($K h^{\prime}[-\Omega \tau+\beta]<0 \Rightarrow \sigma=\rm{Re}(\lambda)\geq 0$), since additional instabilities arise due to time-scale introduced by the filtering process~\cite{Pollakis2014}.
\begin{figure}[t!]
\includegraphics [scale=0.32, angle=0]{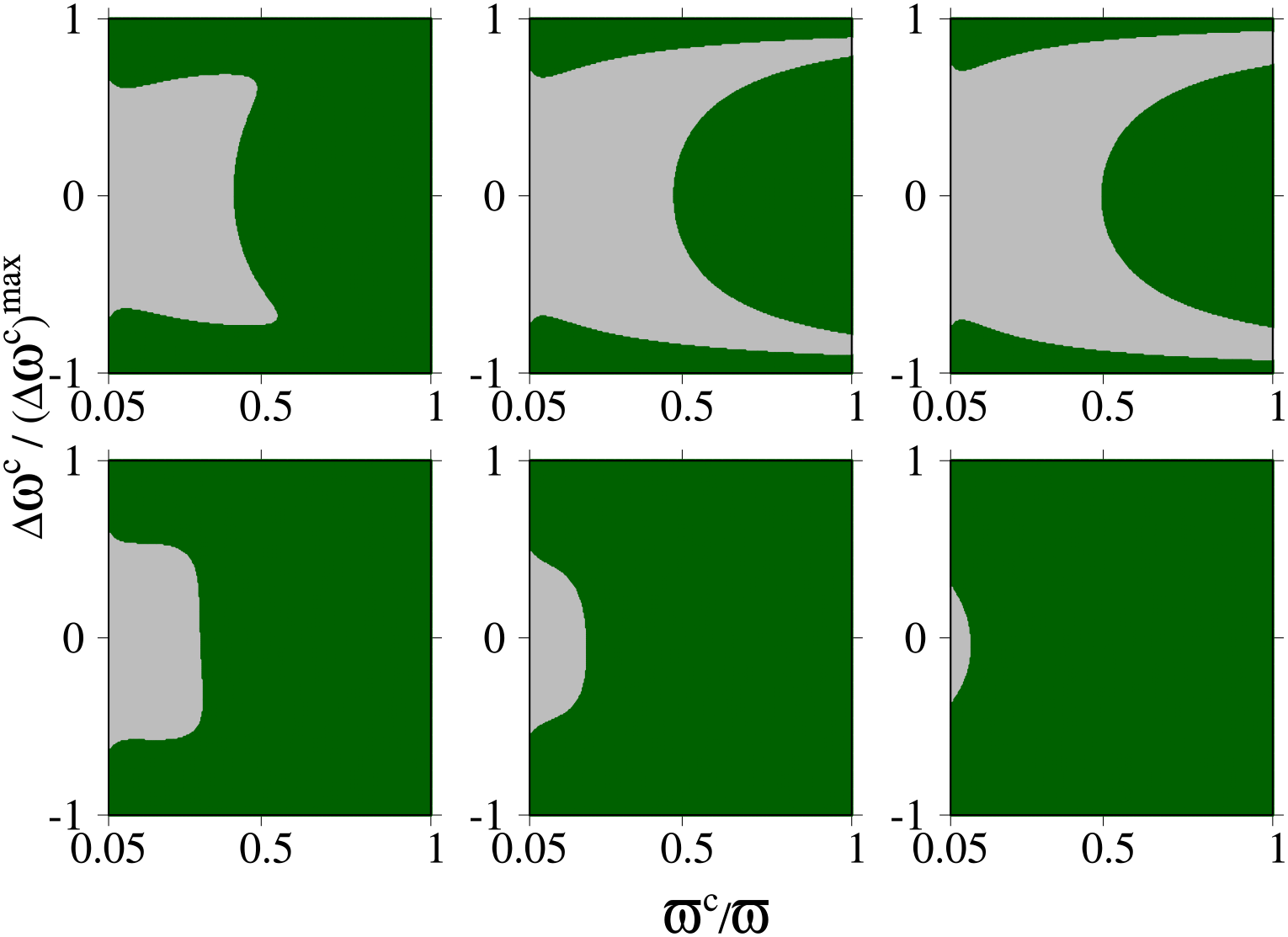}
 \caption{(Color online) 
 Stability of a asymptotic-inphase synchronized state for delay-coupled PLLs with $\Delta\omega = 0.04\times2\pi \rm{~rad Hz}$ plotted as a function of LF parameters -- the mean and the difference of the cut-off frequencies. 
 The green (dark-gray) region denotes the parameter values where the selected solution is stable, and the gray (light gray) the region where it is unstable.
 Results are shown for different transmission-delays (upper row) $\tau=1.1, 1.2, 1.3 \rm{~s}$ at a fixed coupling strength $K=2.0 \rm{~rad Hz}$, and for different coupling strengths (lower row) $K=1.75, 1.5, 1.25 \rm{~rad Hz}$ at fixed delay $\tau=1.1 \rm{~s}$. 
}
\label{fig:new-cof-fig1}
\end{figure}

In Fig.~\ref{fig:new-cof-fig1}, we illustrate the effects of heterogeneous cut-off frequencies on the stability of one branch of the asymptotic-inphase solutions for different values of $K$ and $\tau$.  
The mean cut-off frequency $\bar{\omega}^c = (\omega_{1}^c + \omega_{2}^c)/2$ is varied from a minimum value $0.05\bar{\omega}$ to $\bar{\omega}$.
With the mean $\bar{\omega}^c$ and the difference $\Delta\omega^{c}$, the individual cut-off frequencies of the filters are $\omega_{1,2}^c = \bar{\omega}^c \mp \Delta\omega^{c}/2$. 
The magnitude of difference between cut-off frequencies $\vert \Delta\omega^{c} \vert$ is varied from zero (identical filters) to a maximum $(\Delta\omega^{c})^{\rm{max}} = 2(\bar{\omega}^c - 0.01\bar{\omega})$, such that the lower of the two individual cut-off frequencies (say $\omega_{1}^c = \bar{\omega}^c - \vert \Delta\omega^{c} \vert / 2$) is not smaller than $0.01\bar{\omega}$. 
The normalized difference $\Delta\omega^{c} / (\Delta\omega^{c})^{\rm{max}}$ is plotted on the $y$-axis. 
We find that the stability depends on the mean and the difference between the cut-off frequencies.
For identical LFs ($\omega_{1}^c=\omega_{2}^c$) synchronized states are unstable for cut-off frequencies that are small compared to the mean intrinsic frequency, \ie large integration time of the filters.
However, as heterogeneities are introduced to the LFs, \ie non-identical cut-off frequencies, unstable regimes can become stable.
Hence, specific differences $\Delta\omega^{c}$ in the cut-off frequencies may stabilize synchronized solutions that where unstable in the case of identical LFs. 
The cut-off frequency induced instabilities also depend strongly on the value of the coupling strength $K$ and the transmission-delay $\tau$.
The gray-region indicating unstable solutions becomes considerably smaller as the coupling strength $K$ is decreased. 
This is interesting as in the case without delays and filtering the stability is usually enhanced with larger coupling strength.
The time-scale associated to the loop-filter integration makes the system inert, i.e., effectively responding to outdated information.
As the coupling strength increases, we observe that the instabilities introduced by this inert dynamics are enhanced.
This is related to the perturbation decay dynamics which is underdamped for most values of the transmission delay, see e.g., values of $\gamma$ in Fig. \ref{fig:new-hf-fig11}.
The stronger the VCO reacts to the control-signal, the more it tends to overshoot when close to a synchronized state.

\subsection{Heterogeneity in feedback-delay and interplay between feedback, transmission and processing-delays} \label{subsec:feedback-delay}

So far we have discussed the effects of heterogeneities in the intrinsic frequencies of the VCOs, the transmission-delays and the cut-off frequencies of the LFs. 
The filtering process is associated with a processing time-delay which is inversely proportional to its cut-off frequency $\tau^{\rm c} = b = 1/{\omega^c}$, the LF integration time.
\begin{figure}[t!]
\includegraphics [scale=0.305, angle=0]{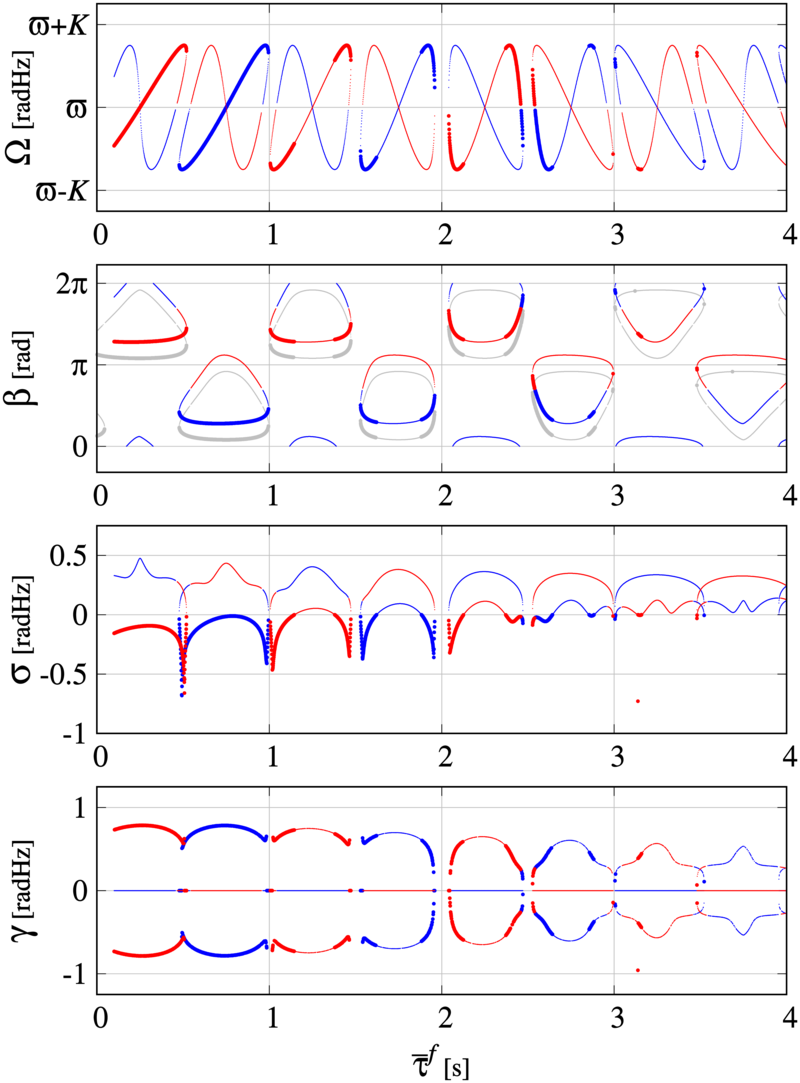}
 \caption{(Color online) 
 Global frequency $\Omega$, phase-difference $\beta$, perturbation response rate $\sigma$ and the corresponding frequency $\gamma$ for two delay-coupled PLLs with heterogeneous intrinsic frequencies $\omega_{1,2} = (1\mp0.02)\times 2\pi \rm{~rad Hz}$ as a function of mean feedback-delay $\bar{\tau}^f$ with a constant feedback-delay difference $\Delta\tau^f=0.2 \rm{~s}$.
 The blue (dark gray) and red (light gray) curves correspond to asymptotic-inphase and -antiphase synchronized states, respectively.
 Thick lines denote stable and the thin lines unstable solutions. 
 The results of synchronized solutions with homogeneous feedback-delays $\Delta\tau^f = 0 \rm{~s}$ are shown in gray. 
 Here, transmission-delays are equal and set to $\tau = 2.0 \rm{~s}$, coupling strength is $K=0.50 \rm{~radHz}$ and the cut-off frequency $\omega_c = 0.25\times \bar{\omega} \rm{~radHz}$. 
}
\label{fig:new-fbd-fig1}
\end{figure}
As these processing time-delays increase, perturbation decay rates become smaller~\cite{Pollakis2014}. 
In this section, we discuss how heterogeneous feedback-delays and its interplay with different delay-time scales (transmission and processing-delays) affects synchronized states.
The frequencies only depend on the mean of the transmission and feedback-delays ($\bar{\tau}$, $\bar{\tau}^f$) while the corresponding phase configurations depend on their differences $\Delta\tau$ and $\Delta\tau^f$. 
The collective effect of these delay-times can be understood in terms of an effective mean delay $\bar{\tau}_{e} = \bar{\tau}-\bar{\tau}^f$ and an effective delay-difference $\Delta\tau_{e} = \Delta\tau -\Delta\tau^f$.
If the feedback-delay is equal to the transmission-delay, $\bar{\tau}_{e}$ will be zero. 
In that case, the dynamical properties of the system change qualitatively, as the multistability induced by the transmission-delay vanishes. 
The solutions are given by
\begin{figure}[t!]
\includegraphics [scale=0.32, angle=0]{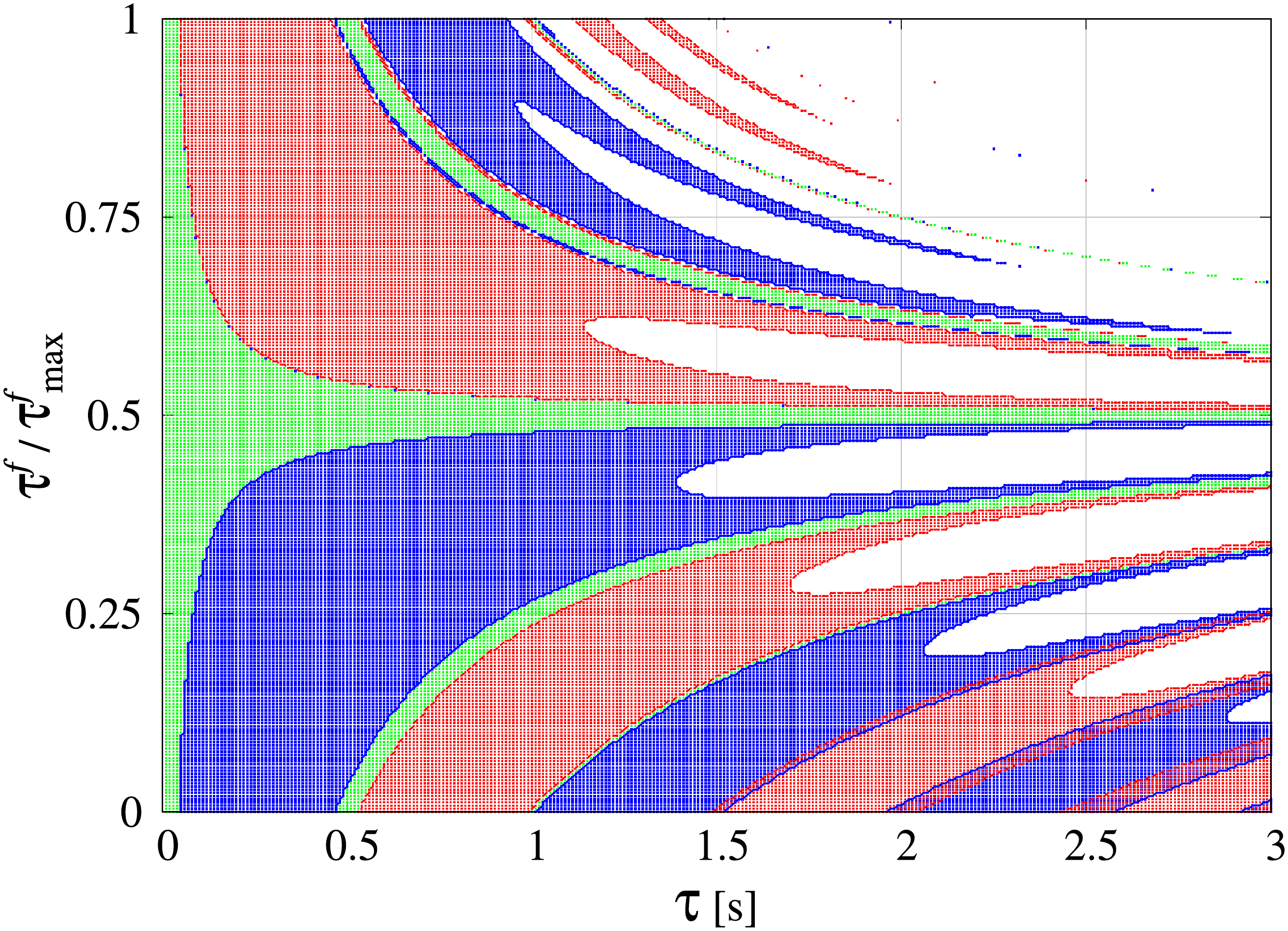}
 \caption{(Color online) 
 The stability of asymptotic-inphase and -antiphase synchronized states for two coupled PLLs in the parameter space of transmission-delay $\tau$ and feedback-delay $\tau_f$. 
 The feedback-delay $\tau^f$ is varied from zero to a maximum $\tau^f_{max} = 2\tau$ and normalized $\tau^f/\tau^f_{max}$.
 The existence and the stability of the synchronized solutions is shown by the different colors, using the color code of Fig.~\ref{fig:new-hf-fig14}. 
 Results shown are for detuned intrinsic frequencies $\Delta \omega = 0.04\times 2\pi \rm{~rad Hz}$, $K=0.50 \rm{~rad Hz}$ and $\omega^c = 0.25\times \bar{\omega} \rm{~rad Hz}$. 
}
\label{fig:new-fbd-fig2}
\end{figure}
\begin{equation}\label{eq:n=2-sfreq-sol-h4}
\begin{split}
\Omega = \bar{\omega} \pm K \cos \left( -\Omega \bar{\tau}_{e} \right) 
\sqrt{1 - 
\left( \frac{\Delta\omega}{2 K \sin \left( \Omega \bar{\tau}_{e} \right)} \right)^2 }, 
\end{split}
\end{equation}
and 
\begin{equation}\label{eq:n=2-sphase-sol-h4}
\beta = 
\begin{cases}
&- \frac{\Omega \Delta\tau_{e}}{2} + 
\sin^{-1} \left( \frac{\Delta\omega}{2 K \sin \left( \Omega \bar{\tau}_{e} \right)} \right) \vspace{0.25cm} \\
&~~~~~~~~~~~~~~\mbox{if}~~ \sin \left( \Omega \bar{\tau}_{e} \right)>0; \\
\\
&- \frac{\Omega \Delta\tau_{e}}{2} +
\pi - \sin^{-1} \left( \frac{\Delta\omega}{2 K \sin \left( \Omega \bar{\tau}_{e} \right)} \right) \vspace{0.25cm} \\ 
&~~~~~~~~~~~~~~\mbox{if}~~ \sin \left( \Omega \bar{\tau}_{e} \right)<0. \\
\end{cases}
\end{equation}
\noindent
In Fig.~\ref{fig:new-fbd-fig1} we plot synchronized solutions and their corresponding eigenvalues as a function of mean feedback-delay $\bar{\tau}^f$, given the individual feedback-delays as $\tau^f_{1,2} = \bar{\tau}^f \mp \Delta\tau^f/2$ with fixed difference $\Delta\tau^f = 0.2\,\rm{s}$. 
\begin{figure}[t!]
\includegraphics [scale=0.32, angle=0]{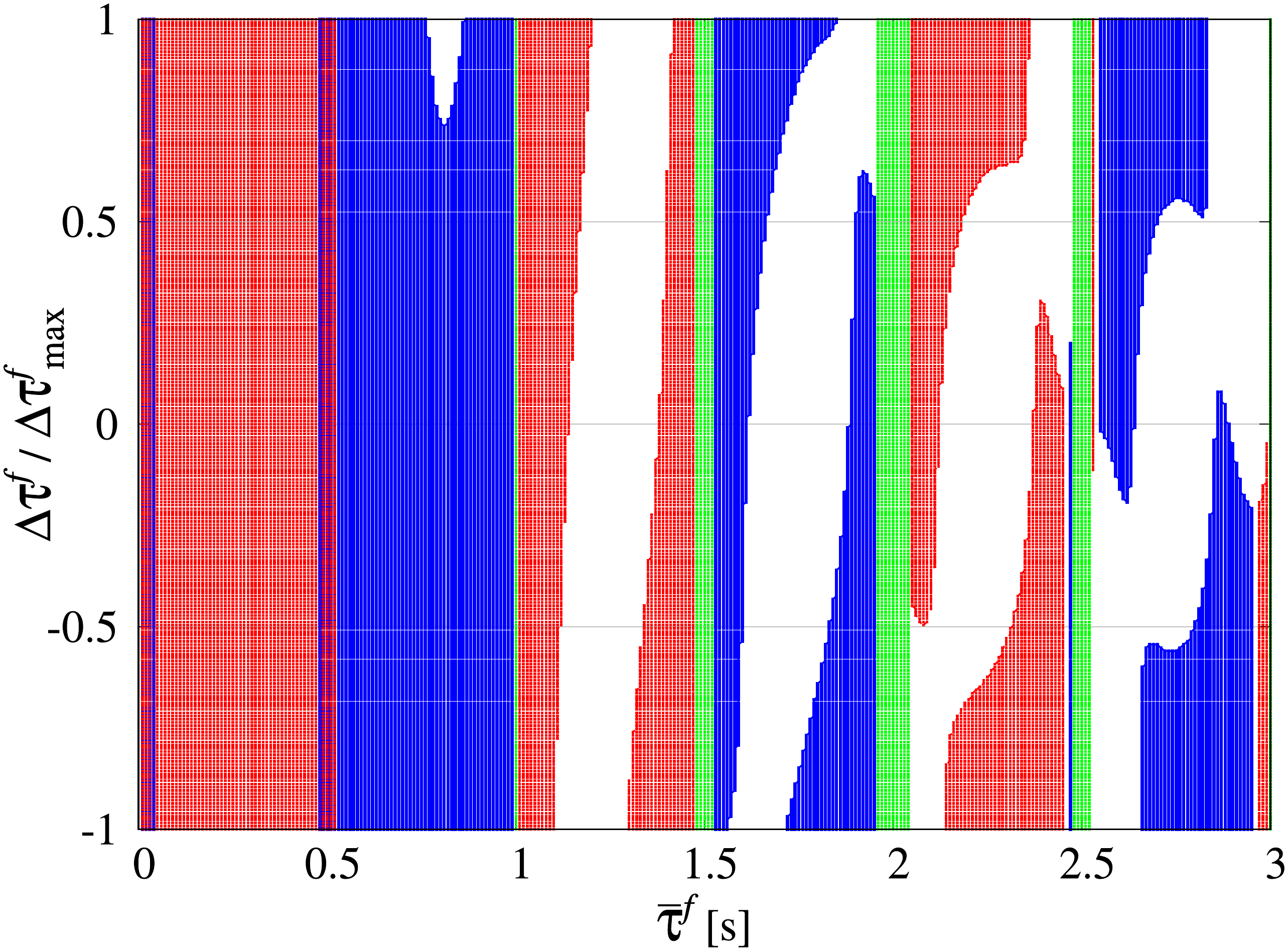}
 \caption{(Color online) 
The stability of asymptotic-inphase and -antiphase synchronized states for two coupled PLLs plotted in the parameter space of the mean feedback-delay $\bar{\tau}^f$ and feedback-delay difference $\Delta\tau^f$.
The existence and the stability of the synchronized solutions is shown by the different colors, using the color code of Fig.~\ref{fig:new-hf-fig14}.
Results shown are for detuned intrinsic frequencies $\Delta \omega = 0.04\times 2\pi \rm{~rad Hz}$, $K=0.50 \rm{~rad Hz}$, $\tau =2 \rm{~s}$, $\omega_c = 0.25\times \bar{\omega} \rm{~rad Hz}$. 
}
\label{fig:new-fbd-fig3}
\end{figure}
We focus on the effects of feedback-delays and therefore consider equal transmission-delays fixed at $\tau = 2\,\rm{s}$. 
In this case there is a symmetry of the frequencies around $\bar{\tau}^{f} = 2\,\rm s$, \ie $\bar{\tau}_{e} = 0$.
%
As can be seen from Eq.~(\ref{eq:n=2-sfreq-sol-h4}) this symmetry around $\bar{\tau}_{e}$ is a generic feature as the cosine and sine squared are symmetric around $0$.
The phase configurations are also modified as in the case of transmission-delays due to the heterogeneity $\Delta\tau^f$, however with opposite sign. 

The characteristic equation~\eqref{eq:n=2-all-hetero-stab-con} for nonzero $\tau^f$ is 
\begin{equation}\label{eq:n=2-stabcon2-h4}
  \begin{split}
  \frac{\lambda^2}{\hat{p}_{1}(\lambda) \hat{p}_{2}(\lambda) e^{-2\lambda \bar{\tau}^{f}}} - 
  \alpha_{12} \alpha_{21} \left( e^{-2\lambda (\bar{\tau} - \bar{\tau}^f)} - 1 \right) +& \\
  \frac{\lambda}{ e^{-\lambda \bar{\tau}^{f}} } 
  \left( 
  \frac{\alpha_{21}}{ \hat{p}_{1}(\lambda) e^{+\lambda \frac{\Delta\tau^{f}}{2} } } + 
  \frac{\alpha_{12}}{ \hat{p}_{2}(\lambda) e^{-\lambda \frac{\Delta\tau^{f}}{2} } } 
  \right)  
  &=0. 
  \end{split}
\end{equation}
  \noindent
For the characteristic equation no single effective delay parameter representing the combined effect of transmission and feedback-delays can be defined. 
%
%
We find that the transmission-delays affect the eigenvalues only through their mean value $\bar{\tau}$.
Fig.~\ref{fig:new-fbd-fig2} and Fig.~\ref{fig:new-fbd-fig3} show that $\Delta\tau^f$ and $\bar{\tau}^f$ have significant effects on the stability of synchronized solutions.
As the feedback-delay is increased, synchronized states become unstable. 
Heterogeneity in the feedback-delays can stabilize and destabilize synchronized solutions that were unstable or stable for identical feedback-delays, respectively.

\subsection{Heterogenous coupling strengths} \label{subsec:coupling}

\begin{figure}[t!]
\includegraphics [scale=0.305, angle=0]{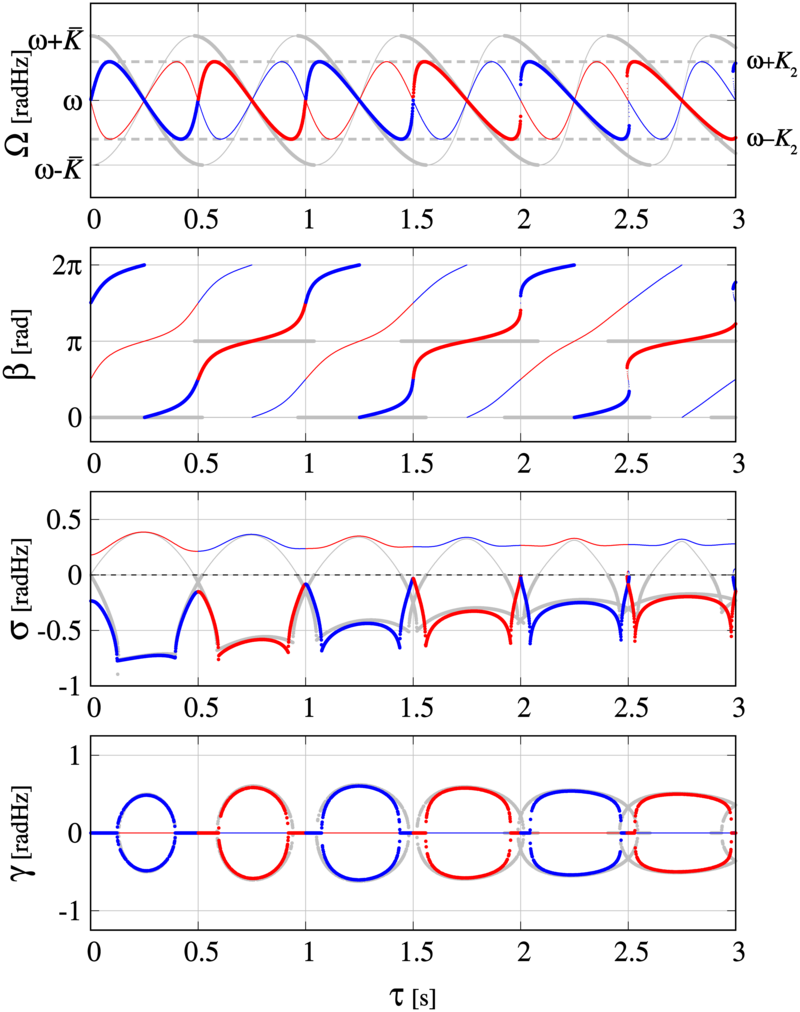}
 \caption{(Color online) 
Global frequency $\Omega$, phase-difference $\beta$, perturbation response rate $\sigma$ and the corresponding frequency $\gamma$ of the synchronized states as a function of the transmission-delay $\tau$ for two mutually delay-coupled PLLs with heterogeneous coupling strengths $\bar{K}=0.25\,\rm{radHz}, \Delta K = -0.2~\rm{radHz}$. 
The intrinsic frequencies are equal $\omega_{1,2} = \omega = 1\times 2\pi~\rm{radHz}$ and the cut-off frequency is $\omega_c=0.25\times\bar{\omega}\,\rm{radHz}$. 
The blue (dark gray) and red (medium gray) curves correspond to the asymptotic-inphase and -antiphase synchronized states.
The thick curves denote stable solutions and the thin curves denote unstable solutions.
The limits of the frequency-range which is accessible to both oscillators with different coupling strength are shown by dashed lines.  
For comparison, the results for equal coupling strengths, \ie for $\Delta K = 0$ are plotted with gray (light gray) curves. 
}
\label{fig:new-hc-fig1}
\end{figure}
\begin{figure}[!th]
\includegraphics [scale=0.305, angle=0]{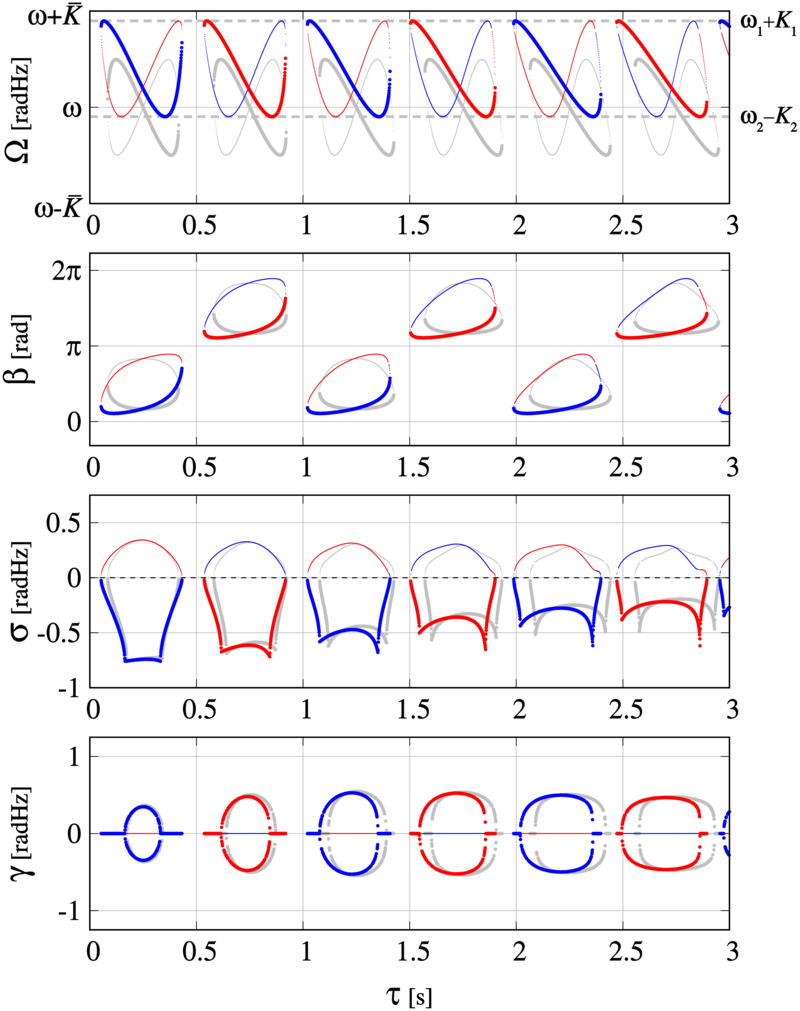}
 \caption{(Color online) 
Global frequency $\Omega$, the phase-difference $\beta$, the perturbation response rate $\sigma$ and the corresponding modulation frequency $\gamma$ of the synchronized states as a function of the transmission-delay $\tau$ for two delay-coupled PLLs with heterogeneous coupling strengths $\bar{K}=0.25\,\rm{radHz}, \Delta K = -0.2~\rm{radHz}$. 
The intrinsic frequencies are $\omega_{1,2} = (1 \mp 0.02) \times 2\pi~\rm{radHz}$ and the cut-off frequency is $\omega_c = 0.25\times\bar{\omega}\,\rm{radHz}$. 
The blue (dark gray) and red (medium gray) curves correspond to the asymptotic-inphase and -antiphase synchronized states.
The thick curves denote stable solutions and the thin curves denote unstable solutions.
The limits of the frequency-range which is accessible to both oscillators with different coupling strength are shown by dashed lines.  
For comparison, the results for equal coupling strengths, \ie for $\Delta K = 0$ are plotted with gray (light gray) curves. 
}
\label{fig:new-hc-fig2}
\end{figure}
\begin{figure}[!th]
\includegraphics [scale=0.305, angle=0]{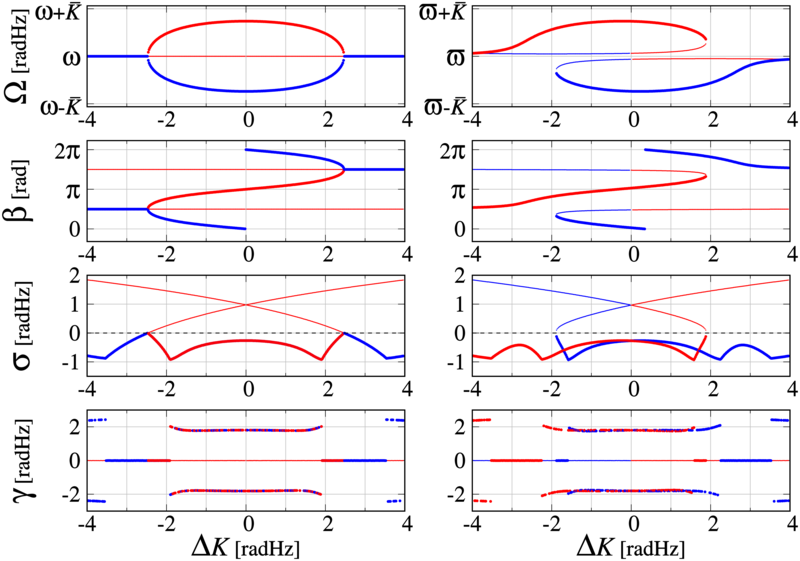}
 \caption{(Color online) 
Global frequency $\Omega$, phase-difference $\beta$, perturbation response rate $\sigma$ and the corresponding frequency $\gamma$ of the synchronized states as a function of difference in coupling strength $\Delta K = (K_2 - K_1)$ for two delay-coupled PLLs at fixed mean coupling $\bar{K}=2.0\,\rm{radHz}$, transmission-delay $\tau = 0.5~\rm{s}$ and cut-off frequency $\omega_c=0.25\times\bar{\omega}\,\rm{radHz}$. 
The blue (dark gray) and red (light gray) curves correspond to the inphase and antiphase ($\Delta\omega = 0\,\rm{radHz}$) or asymptotic-inphase and -antiphase ($\Delta\omega = 0.04\times 2 \pi\,\rm{radHz}$) synchronized states.
The thick curves denote stable solutions (from Eq.~\eqref{eq:n=2-all-hetero-stab-con}, with $\sigma=\Re(\lambda_{max})<0$) and the thin curves unstable solutions.
The left column shows the results for $\omega_{1,2} = 1 \times 2\pi\,\rm{~rad Hz}$ and the right column the results for $\omega_{1,2} = (1 \mp 0.02) \times 2\pi\,\rm{radHz}$.
}
\label{fig:new-hc-fig3}
\end{figure}
The coupling strength $K$ denotes the sensitivity of the VCO to the control signal.
It interacts with the effects of the transmission- and feedback-delays as well as the filtering process on the dynamics of the coupled system.
For example from Ref.~\cite{Schuster1989} it is well-known that the effects of time-delays are more pronounced for larger coupling strengths, see e.g., Fig.~\ref{fig:new-hf-dd-fig1}, and in the Supplementary material Figs.~\ref{fig:new-hf-fig12}-\ref{fig:new-hc-fig4}, for how the number of synchronized states increases with larger coupling $K$. 
Simultaneously, more synchronized solutions become unstable as the instabilities introduced by the filtering process become more prominent. 
On the other hand, the coupling above the critical coupling strength enables synchrony in detuned oscillators by forcing them towards a common frequency, and therefore counters the effects of detuning $\Delta\omega$. 
These effects are modified when PLLs with heterogeneous sensitivity interact with each other, \ie their coupling strengths are heterogeneous, see Eq.~\eqref{eq:n=2-all-hetero-sync-sol}. 
Note that for identical coupling strengths and intrinsic frequencies, there are non-generic solution for specific parameter values when $\sin(\Omega \bar{\tau}_e)=0$. 
These non-generic pairs of solutions with coinciding frequencies split-up as a result of the symmetry breaking, i.e., $K_1 \neq K_2$ and the system does not become underdetermined. 
For such heterogeneity in the coupling strengths, the phase-differences and frequencies of the synchronized states are significantly modified, see Fig.~\ref{fig:new-hc-fig1} and~\ref{fig:new-hc-fig2}.
For $\Delta\omega = 0$, the maximum deviation $\vert\Omega-\omega\vert$ is now determined by the smallest of the coupling strengths, e.g., $\Omega \in [\omega - K_2, \omega + K_2]$ in Fig.~\ref{fig:new-hc-fig1}. 
For detuned PLLs the range of global frequencies is given by $\Omega \in [\hbox{max}(\omega_k - K_k), \hbox{min}(\omega_k + K_k)]$, where $k = 1,2$.
The stability of these solutions and is given by Eq.~\eqref{eq:n=2-all-hetero-stab-con}.
The perturbation decay rate depends non-linearly on the difference between the coupling strengths, see Fig.~\ref{fig:new-hc-fig3}. 
For differences in the coupling strength where one of them approaches zero and the other $\bar{K}$, the decay rates approach the case of perturbation decay in entrained oscillator systems.
However, the maximum decay rates are found for values of $\Delta K$ that do not equal to the case of entrainment $\Delta K_{\rm max}$ or identical coupling, $\Delta K=0$.
For these values where perturbations decay fastest, the decay changes qualitatively from oscillatory to overdamped.
This happens as the change in $\Delta K$ causes different solutions of the characteristic equation with zero and non-zero imaginary parts to interchange in terms of the maximum real part $\sigma$.
Fig.~\ref{fig:new-hc-fig2} suggests that around $\Delta K=\mp 2$ the system undergoes a pitchfork-bifurcation and reverse.
As soon as detuning of the PLLs is introduced, this changes and something closer to a transcritical bifurcations arises and frequencies of the different synchronized states change gradually with $\Delta K$ until they crash onto the other branch.
That means that a slight detuning of the intrinsic frequencies of the PLLs can cause an abrupt change in the frequencies of synchronized states.

\section{Experiments and Spice simulations} \label{sec:exp-sim}

\begin{figure}[b!]
\includegraphics [scale=0.41, angle=0]{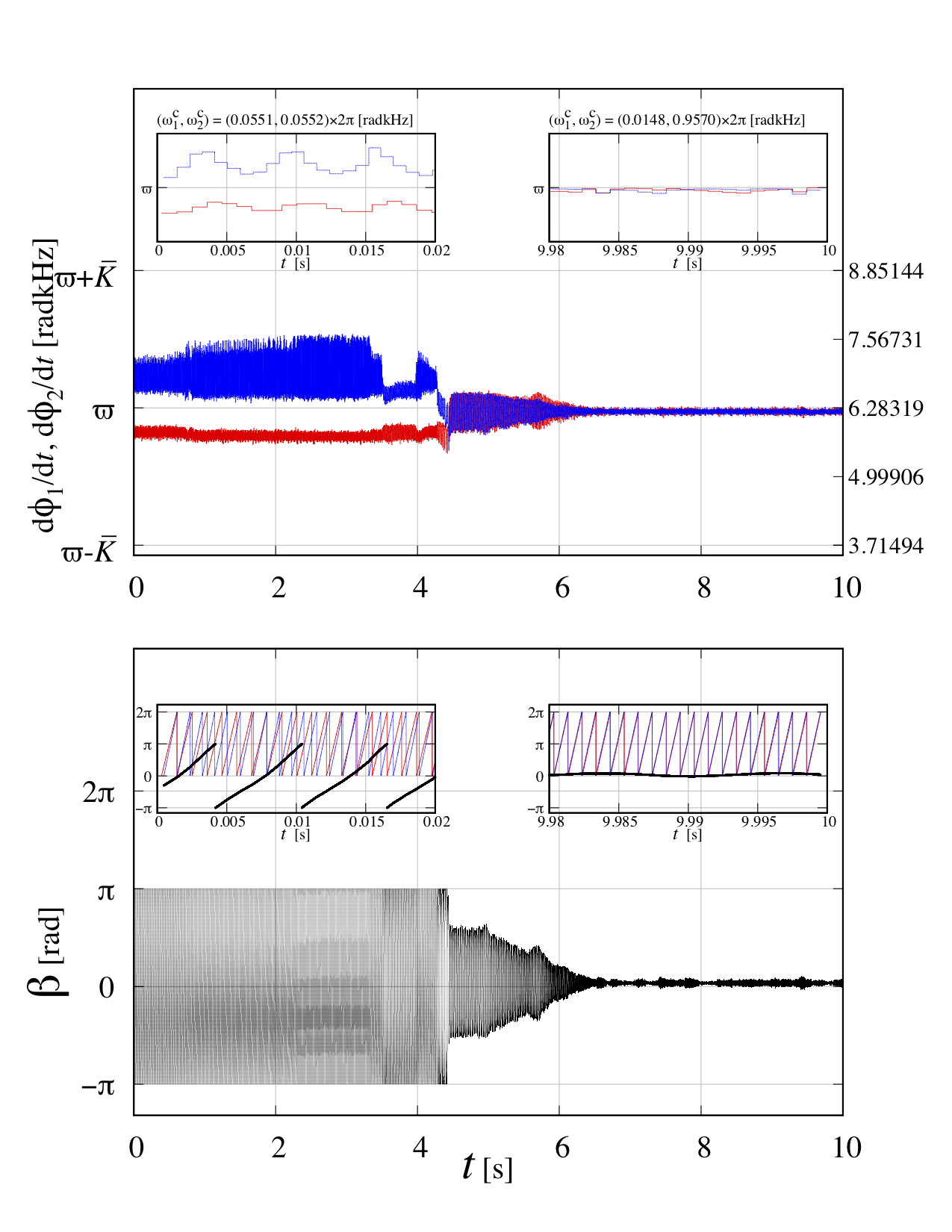}
 \caption{(Color online) 
  Experimental realization of stabilization of a synchronized state by introducing heterogeneity in the cut-off frequencies of the loop-filters $\omega_{1,2}^c$. 
  Initially the cut-off frequencies are equal $\omega_{1,2}^c=(0.055, 0.055)\times 2\pi \, \rm{radkHz}$, then are changed to $\omega_{1,2}^c=(0.0148, 0.9570)\times 2\pi \, \rm{radkHz}$ during the experiment. 
  It is observed that a asymptotic-inphase state stabilizes as a result of introducing heterogeneity in $\omega_{1,2}^c$. 
  The transmission delay values are equal and set to $\tau = 0.7512 \, \rm{ms}$. 
  The insets show the frequencies, phases (blue, red) and phase differences (black) obtained from the measured signals.
}
\label{fig:new-exp-dwc}
\end{figure}
To validate our theoretical predictions, we performed experiments using prototype DPLLs.
These are coupled via a microcontroller that buffers and thereby delays their output signals~\cite{Wetzel2017}.
The DPLL parameters are given in Table~\ref{tab:DPLLparam}.
In the setup currently available we can tune the transmission delay via the microcontroller and the cut-off frequency using the tunable resistance of the RC loop-filter.   
The coupling-strength and the intrinsic frequencies are fixed system properties whose heterogeneity has been measured and is known~\cite{Wetzel2017}.
\begin{table}[th]
\begin{tabular}{|c|c|c|c|}
\hline
	& $\omega$ in Hz & $K$ in Hz/V & $\omega^c$ in Hz \\ \hline
 DPLL 1 & 948  & 409.5 & 14.8 				  \\ \hline
 DPLL 2 & 1008 & 408   & 15.4 				\\ \hline
 \end{tabular}%
\caption{Prototype DPLL parameters.}
\label{tab:DPLLparam}
\end{table}

Feedback-delays, inverters and dividers are not available in this prototype setup.
A detailed description of how the experiments were carried out can be found in the Supplementary material, section~\ref{supp-mat-exp-spice}. 

We also performed circuit-level \textit{Spice} simulations using the freeware implementation LTspice~\cite{Nagel1973,LTspice}. 
In this framework, the voltages, currents and delay-times associated with the components of the electronic architecture can be simulated, see Fig.~\ref{fig:sketchLTspiceCircuit} in the Supplementary material.
The electronic components in the \textit{Spice} simulation can be configured to match those of the experimental setup by tuning individual voltages, resistances, propagation-delays and capacitances.

\begin{figure}[t!]
\includegraphics [scale=0.305, angle=0]{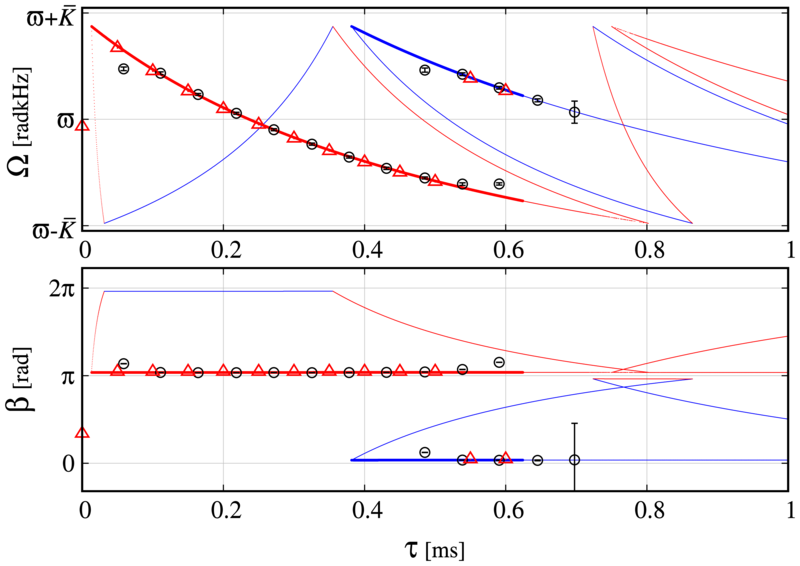}
 \caption{(Color online) 
  Experimental results for global frequency $\Omega$ and phase-difference $\beta$ as a function of transmission delay $\tau$ for two heterogeneous delay-coupled digital PLLs. 
  Experimental results from prototype PLLs and simulations results from LTspice are plotted with circles and triangles respectively, analytic results are represented by the curves. 
}
\label{fig:new-exp-tau-fig11}
\end{figure}

\begin{figure}[h!]
\includegraphics [scale=0.305, angle=0]{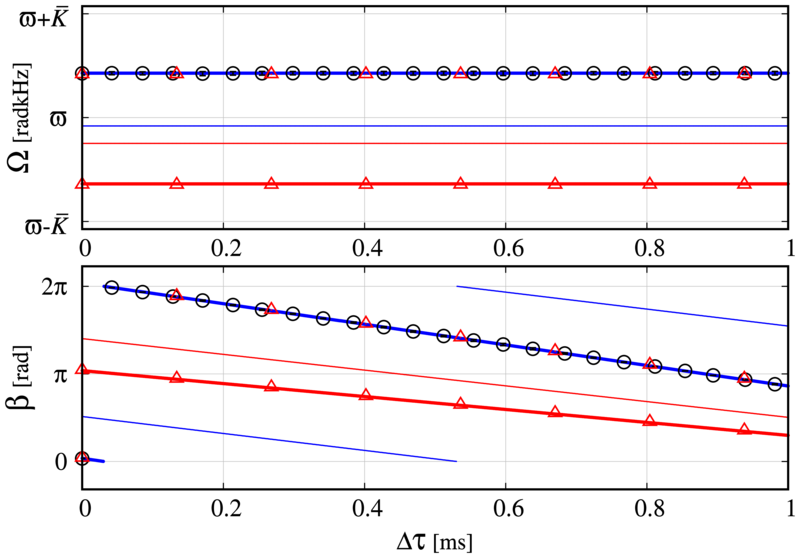}
 \caption{(Color online) 
  Experimental results for global frequency $\Omega$ and phase-difference $\beta$ as a function of transmission delay difference $\Delta\tau$ for two heterogeneous delay-coupled digital PLLs. 
  The results are obtained at fixed mean delay value of $\bar{\tau} = 0.536 \, \rm{ms}$.
  Experimental results from prototype PLLs and simulations results from LTspice are plotted with circles and triangles respectively, whereas the analytic results are given by the curves. 
}
\label{fig:new-exp-dtau-fig12}
\end{figure}

Taking into account the measured heterogeneity in the intrinsic frequencies and coupling strengths of the DPLL prototypes we show in theory, experiment and simulation that the phase-model can precisely predict the phase-differences and global frequencies of synchronized states, see Fig.~\ref{fig:new-exp-tau-fig11}.
We also show how the phase-difference depends linearly on the difference in the delays, while the frequency of the synchronized state remains constant as predicted in subsection~\ref{subsec:transmission-delay}, see Fig.~\ref{fig:new-exp-dtau-fig12}.
Furthermore we show an example, how synchrony can be recovered for an unstable synchronized state by tuning the cut-off frequencies to heterogeneous values while keeping the mean cut-off frequency constant, see Fig.~\ref{fig:new-exp-dwc}.
This is achieved by tuning the resistance trimmers of two coupled PLLs at runtime from identical values to values that correspond to detuned cut-off frequencies while the mean cut-off frequency is kept constant.
As a result the synchronized state becomes stable.


\section{Discussion and application} \label{sec:discuss}

In this work we studied how breaking parameter symmetry affects the synchronization dynamics in  systems of mutually delay-coupled oscillators.
We focused our analysis on the tractable case of two oscillators.
For studying larger systems we provide the analytic expressions, but their analysis is beyond the scope of this paper.
How the effects found for the two PLL system change with, e.g., coupling topologies, the number of oscillators, or in dependence of the distributions of the heterogeneities are interesting starting points for further research and need to be addressed when moving into application.
Certainly the control of self-organized synchronization will be highly relevant for the design of large distributed networks of mutually delay-coupled clocks.
The self-organization approach to synchronization discussed here is fundamentally and qualitatively different from hierarchical approaches.
There is no accumulation of phase-errors between the oscillators and signaling delays do not necessarily cause phase-offsets between the oscillators in synchronized states, to name two key strengths of self-organized synchronization.
However, the complex dependencies of the dynamics on the system parameters pose a challenge when it comes to the control and robustness of synchronization in the light of process, voltage and temperature (pvt)-variations, relevant to modern microelectronics. 
Contrary to the expected hindering effects of component heterogeneity, known from synchronization in hierarchical networks, we here showed that clock heterogeneities can actually enhance synchronization in networks with flat hierarchy. 
Specifically, this reflects in the optimal difference of the coupling-strengths that optimizes the decay of perturbations to synchronized states, and heterogeneous LF cut-off frequencies that can stabilize synchronized states.
Furthermore time-delays that usually complicate the synchronization of spatially distributed clocks are essential for the stability of synchronized states and control the clocks' mutual phase-differences.
Interestingly this dependence of the phase-configurations on the effective delays is linear.

Hence, the complex interplay between the different system parameters leads to rich dynamics and can provide the means to tune into synchronized states with, e.g., arbitrary phase relations.
These findings have strong implications for the application of self-organized synchronization in engineering. 
Such an approach without hierarchical structures can only be taken if synchronization dynamics can be controlled.  
As heterogeneities in the clock components are inevitable in modern integrated circuitry, adding transmission and feedback-delay elements and allowing for tunable loop-filters can enhance the control of the dynamics and perturbation decay processes in such systems.
Defined phase differences between frequency synchronized clocks for example enable \textit{beamforming}, i.e., spatially focused emission of electro-magnetic waves, relevant for communication, audio and radar applications \cite{Agrawal1999,Hajimiri2005,Theodoropoulos2009,Hu2013}.
A common, spatially distributed time-reference with locked phases can be the enabler of precise indoor-navigation systems, as the accuracy of the localization depends directly on the synchrony of the satellite-nodes.
Servers for globally available data-bases could be kept in sync to reduce the time-uncertainty of their time-stamps~\cite{Corbett2012, Brewer2017}.
Furthermore, cut-off frequencies much smaller than the mean intrinsic frequency make such clock networks inert and lead to slow perturbation decay~\cite{Pollakis2014}.
At the same time however, such small cut-off frequencies will enhance the signal-to-noise ratio within the circuitry~\cite{Widrow1975}.
These properties, together with the results that heterogeneity in the cut-off frequency can stabilize synchronized states that would have otherwise been unstable, show the potential for optimization of synchronization in large networks of spatially distributed electronic clocks~\cite{Lee2000,Lim2000}.
We envision that the effects studied here will be used in next-generation synchronization layers for large spatially distributed systems.
All free system parameters that can be tuned as the system operates would then be used to steer self-organized synchronization dynamics towards application-specific requirements within the same conceptual setting.

\section{Acknowledgments}
We thank Johannes Fritzsche, Frank J\"{u}licher, David J\"{o}rg, Rabea Seyboldt, Kevin Bassler, and Lennard Hilbert for stimulating discussions. This work is partly supported by the German Research Foundation (DFG) within the Cluster of Excellence `Center for Advancing Electronics Dresden'.

\begin{widetext}

\appendix

\section{Synchronized solutions} \label{sec:app-sync-sol-N2}

In the following, we obtain analytic expressions for global frequencies and phase differences in synchronized states for two delay-coupled PLLs with heterogeneous parameters. 
We consider cosine coupling function, for which the phase dynamics is given by Eq.~\eqref{eq:n=2-system} as, 
\begin{equation}\label{eq:app-n=2-system}
\begin{split}	
	\dot{\phi}_{i}(t) = \omega_{i} + 
	K_{i} \int_0^{\infty} \du p_{i}(u) \cos \left( \phi_{j} (t-u-\tau_{ij}) - \phi_{i} (t - u -\tau^{f}_{i}) \right).
\end{split}
\end{equation}
Here the indices $i,j=1,2$; $j\neq i$; we will use this convention for $i,j$ here on unless specified otherwise. We are interested in phase-locked synchronized states, i.e., when the oscillators evolve with same collective frequency $\Omega$ and have a constant phase-lag $\beta$ between them, which leads to the ansatz
\begin{equation}\label{eq:app-n=2-sync-ansatz}
\begin{split}	
	\phi_1 = \Omega t;~~\phi_2 = \Omega t + \beta.
\end{split}
\end{equation}
Using this ansatz in Eq.~\eqref{eq:app-n=2-system} one has
\begin{equation}\label{eq:app-n=2-sub-solutions-gen}
\begin{split}
\Omega &= 
\omega_{i} + K_{i} \cos \left( -\Omega (\tau_{ij} - \tau^f_{i}) \pm \beta \right).
\end{split}
\end{equation}
Rewriting this equation in terms of mean-coupling $\bar{K}$ and coupling-difference $\Delta K$, this equation reads, 
\begin{equation}\label{eq:app-n=2-hcpl-eq1}
\begin{split}
\Omega &= 
\omega_{i} + (\bar{K} \mp \Delta K/2) \cos \left( -\Omega (\tau_{ij} - \tau^f_{i}) \pm \beta \right).
\end{split}
\end{equation}
Subtracting these equations, we have 
\begin{equation}\label{eq:app-n=2-hcpl-eq2}
\begin{split}
2 \bar{K} \sin(\Omega \bar{\tau}_e) \sin(B) - \Delta K \cos(\Omega \bar{\tau}_e) \cos(B) = \Delta \omega, 
\end{split}
\end{equation}
where $B = (\Omega \Delta{\tau_e})/2 + \beta$. As defined previously, $\bar{\tau}_e = (\bar{\tau} - \bar{\tau}^f)$ and $\Delta{\tau_e} = (\Delta \tau - \Delta \tau^f)$ are respectively the effective mean delay and the effective delay-difference. 
Dividing Eq.~\eqref{eq:app-n=2-hcpl-eq2} by $H$ defined as 
\begin{equation}\label{eq:app-n=2-hcpl-eq3}
\begin{split}
H = \sqrt{ \left(2 \bar{K} \sin(\Omega \bar{\tau}_e) \right)^2 + \left( \Delta{K} \cos(\Omega \bar{\tau}_e) \right)^2 },
\end{split}
\end{equation}
provides, 
\begin{equation}\label{eq:app-n=2-hcpl-eq4}
\begin{split}
B = \sin^{-1}\left( \frac{\Delta\omega}{H}\right) + 
\sin^{-1}\left( \frac{\Delta{K} \cos(\Omega \bar{\tau}_e)}{H}\right).
\end{split}
\end{equation}
Therefore, the phase difference $\beta$ in the synchronized state is given by, 
\begin{equation}\label{eq:app-n=2-hcpl-sphase}
\begin{split}
\beta = - \frac{\Omega \Delta \tau_e}{2} + \sin^{-1}\left( \frac{\Delta\omega}{H}\right) + 
\sin^{-1}\left( \frac{\Delta{K} \cos(\Omega \bar{\tau}_e)}{H}\right).  
\end{split}
\end{equation}
We rewrite this expression as
\begin{equation}\label{eq:app-n=2-hcpl-sphase}
\begin{split}
\beta = - \frac{\Omega \Delta \tau_e}{2} + \sin^{-1}\left( \tilde{A}\sqrt{1-\tilde{B}^2}+\tilde{B}\sqrt{1-\tilde{A}^2} \right),  
\end{split}
\end{equation}
where $\tilde{A}=\Delta\omega/H$ and $\tilde{B}=\Delta{K} \cos(\Omega \bar{\tau}_e)/H$.
Depending on positive and negative values of $\bar{K} \sin(\Omega \bar{\tau}_e)$ and $\Delta{K} \cos(\Omega \bar{\tau}_e)$, above expression provide asymptotic-inphase and -antiphase configurations in the synchronized state. 
\begin{equation}\label{eq:app-n=2-sphase-sol-h1}
\beta = 
\begin{cases}
- \frac{\Omega \Delta \tau_e}{2} + \left[ \sin^{-1}\left( \frac{\Delta\omega}{H}\right) + 
\sin^{-1}\left( \frac{\Delta{K} \cos(\Omega \bar{\tau}_e)}{H}\right) \right],
~~\mbox{if}~~\bar{K}\sin \left( \Omega {\tau_e} \right)>0,~\rm{(asymptotic-inphase)}\\
\\
- \frac{\Omega \Delta \tau_e}{2} + 2\pi - \left[ \sin^{-1}\left( \frac{\Delta\omega}{H}\right) + 
\sin^{-1}\left( \frac{\Delta{K} \cos(\Omega \bar{\tau}_e)}{H}\right) \right],
~~\mbox{if}~~\bar{K}\sin \left( \Omega {\tau_e}  \right)<0,~\rm{(asymptotic-inphase)}\\
\\
- \frac{\Omega \Delta \tau_e}{2} + \pi - \left[ \sin^{-1}\left( \frac{\Delta\omega}{H}\right) - 
\sin^{-1}\left( \frac{\Delta{K} \cos(\Omega \bar{\tau}_e)}{H}\right) \right],
~~\mbox{if}~~\bar{K}\sin \left( \Omega {\tau_e}  \right)>0,~\rm{(asymptotic-antiphase)}\\
\\
- \frac{\Omega \Delta \tau_e}{2} + \pi + \left[ \sin^{-1}\left( \frac{\Delta\omega}{H}\right) - 
\sin^{-1}\left( \frac{\Delta{K} \cos(\Omega \bar{\tau}_e)}{H}\right) \right],
~~\mbox{if}~~\bar{K}\sin \left( \Omega {\tau_e}  \right)<0,~\rm{(asymptotic-antiphase)}\\
\end{cases}
\end{equation}

Adding Eqs.~\eqref{eq:app-n=2-hcpl-eq1}, provide the expression for collective frequencies as, 
\begin{equation}\label{eq:app-n=2-hcpl-sfr}
\begin{split}
\Omega = \bar{\omega} + \bar{K} \cos(\Omega \bar{\tau}_e) \cos(B) - \frac{\Delta K}{2} \sin(\Omega \bar{\tau}_e) \sin(B). 
\end{split}
\end{equation}
One can obtain the frequencies of synchronized state from above equation by substituting the value of $B$ from Eq.~\eqref{eq:app-n=2-hcpl-eq4}.

In sec. \ref{sec:results}, equal coupling strengths are assumed to study the effects of heterogeneous intrinsic frequencies, filters and different delay-times. 
For this case ($K_1 = K_2 = K$), using Eq.~\eqref{eq:app-n=2-sub-solutions-gen}, the expressions for synchronized solutions can be obtained as following. 
\begin{equation}\label{eq:app-n=2-sub-solutions}
\begin{split}
\Omega &= 
\omega_i + K \cos \left( -\Omega (\tau_{ij} - \tau^f_{i}) \pm \beta \right).
\end{split}
\end{equation}
Using Eq.~\eqref{eq:app-n=2-hcpl-eq2} and \eqref{eq:app-n=2-hcpl-sfr} for $\Delta K = 0$ yields
\begin{equation}\label{eq:app-n=2-sph2}
\begin{split}
\sin \left( \Omega( \bar{\tau} - \bar{\tau}^{f} ) \right) \sin \left( \frac{\Omega(\Delta\tau - \Delta\tau^{f})}{2} + \beta \right) = \frac{\Delta\omega}{2 K},
\end{split}
\end{equation}
\noindent
and 
\begin{equation}\label{eq:app-n=2-sfr1}
\begin{split}
\Omega = \bar{\omega} + K \cos \left( -\Omega (\bar{\tau} - \bar{\tau}^{f}) \right) 
\cos \left( \frac{\Omega(\Delta\tau - \Delta \tau^{f})}{2} + \beta \right). 
\end{split}
\end{equation}
\noindent

For oscillators with identical frequencies ($\omega_1 = \omega_2$) and delay-values such that $\Omega ( \bar{\tau} - \bar{\tau}^{f} ) = n \pi,~ n= 1,2,3 \cdots$, pair of non-generic solutions exist with equal global frequencies and two distinct phase difference values. 
Consider Eq.~\eqref{eq:app-n=2-sub-solutions} for identical oscillators $\omega_1 = \omega_2$, coupled with symmetric delay values $\tau_{12} = \tau_{21}$ and $\tau^f_1 = \tau^f_2$. When $\Omega ( {\tau}_{ij} - {\tau}_{i}^{f} ) = n \pi$, 
$$\Omega = \omega + K \cos(n \pi - \beta) = \omega + K \cos(n \pi + \beta).$$
Multiplying by $(\tau-\tau^f)$ and using $\Omega (\tau-\tau^f) = n\pi$, we have 
$$ n \pi - \omega (\tau-\tau^f) = K (\tau-\tau^f) \cos(n \pi - \beta) = K (\tau-\tau^f) \cos(n \pi + \beta);$$
$$\frac{n \pi - \omega (\tau-\tau^f)}{K (\tau-\tau^f) } = \cos(n \pi - \beta) = \cos(n \pi + \beta).$$ 
Due to symmetry in the coupling function, above criteria is fulfilled for two distinct values of $\beta$ with coinciding global frequency. This degeneracy is lost and the solutions split up when symmetry is broken due to parameters heterogeneity. 

For generic solutions, we require $\sin \left( \Omega( \bar{\tau} - \bar{\tau}^{f} ) \right) \neq 0$ and obtain the phase differences $\beta$ that are allowed in the synchronized states. 
From Eq.~\eqref{eq:app-n=2-sph2} the phase difference $\beta$ is 
\begin{equation}\label{eq:app-n=2-sphase-sol}
\beta = 
\begin{cases}
&- \frac{\Omega(\Delta\tau - \Delta\tau^{f})}{2} + 
\sin^{-1} \left( \frac{\Delta\omega}{2 K \sin \left( \Omega( \bar{\tau} - \bar{\tau}^{f} ) \right)} \right) \vspace{0.25cm}\\
&~~~~~~~~~~~~\mbox{if}~~\sin \left( \Omega( \bar{\tau} - \bar{\tau}^{f} ) \right)>0, \\
\\
&- \frac{\Omega(\Delta\tau - \Delta\tau^{f})}{2} +
\pi - \sin^{-1} \left( \frac{\Delta\omega}{2 K \sin \left( \Omega( \bar{\tau} - \bar{\tau}^{f} ) \right)} \right) \vspace{0.25cm}\\
&~~~~~~~~~~~~~\mbox{if}~~\sin \left( \Omega( \bar{\tau} - \bar{\tau}^{f} ) \right)<0. \\
\end{cases}
\end{equation}
\noindent
A transcendental equation for the collective frequency as a function of the phase shift $\beta$, the mean delays $\bar{\tau},\bar{\tau}^f$, the mean intrinsic frequency $\bar{\omega}$ and the detuning $\Delta\omega$ 
can then be obtained from Eq.~\eqref{eq:app-n=2-sphase-sol} and Eq.~\eqref{eq:app-n=2-sfr1}

\begin{equation}\label{eq:app-n=2-sfreq-sol}
\begin{split}
\Omega = \bar{\omega} \pm K \cos \left( -\Omega (\bar{\tau} - \bar{\tau}^{f}) \right) 
\sqrt{1 - 
\left( \frac{\Delta\omega}{2 K \sin \left( \Omega( \bar{\tau} - \bar{\tau}^{f} ) \right)} \right)^2 }.
\end{split}
\end{equation}
The generalization of this analysis to finite systems with $N$ oscillators and arbitrary $2\pi$-periodic coupling functions is provided in the Supplementary material, see \ref{supp-mat-Nosc}. 

\section{Stability of the synchronized solutions} \label{sec:app-stab} 

\noindent
In this section we study the linear stability of the self-organized synchronized states characterized in the previous section.
These solutions to Eq.~\eqref{eq:app-n=2-system} are characterized by common collective frequencies and constant phase differences between the oscillators. 
Modifying Eq.~\eqref{eq:app-n=2-sync-ansatz}, we ask how the system responds if subject to small perturbations $q_{1,2}(t)$
\begin{equation}\label{eq:app-n=2-small-pert}
\phi_1(t) = \Omega t + q_1(t), 
~ \phi_2(t) = \Omega t + \beta + q_2(t). 
\end{equation}
\noindent
With this ansatz and using Eq.~\eqref{eq:app-n=2-system} we obtain the dynamics of weakly perturbed synchronized states   
\begin{equation}\label{eq:app-n=2-pert01}
\begin{split}
\Omega + \dot{q}_{i}(t) 
= \omega_{i} + K  \int_0^{\infty} \du p_{i}(u) 
~h \left[ 
\left( -\Omega(\tau_{ij}-\tau^{f}_{i}) \pm \beta \right) 
+ \left( q_{j}^{u+\tau_{ij}} - q_{i}^{u+\tau^{f}_{i}} \right) 
\right].
\end{split}
\end{equation}
\noindent
Here we introduced the notations $q_{i}^{u+\tau_{ji}} = q_{i}(t-u-\tau_{ji})$ and $q_{i}^{u+\tau^{f}_{i}} = q_{i}(t-u-\tau^{f}_{i})$. 
Expanding the coupling function to first order, the perturbation dynamics can be separated  
\begin{equation}\label{eq:app-n=2-pert02}
\begin{split}
\dot{q}_{i}(t) = K \int_0^{\infty} \du p_{i}(u) ~h^{\prime} \left[ \left( -\Omega(\tau_{ij}-\tau^{f}_{i}) \pm \beta \right) \right] \left( q_{j}^{u+\tau_{ij}} - q_{i}^{u+\tau^{f}_{i}} \right).
\end{split}
\end{equation}
\noindent
Then, using the exponential ansatz, \ie $q_{i}(t) = d_{i} \exp(\lambda t)$ we obtain
\begin{equation}\label{eq:app-pert04}
\begin{split}
\lambda ~d_i = \hat{p}_{i}(\lambda) e^{-\lambda \tau^{f}_{i}} ~\alpha_{ij}~  
\left( d_{j} e^{-\lambda (\tau_{ij}-\tau^{f}_{i})} - d_{i} \right), 
\end{split}
\end{equation}
where $\alpha_{ij} = K ~h^{\prime} [ ( -\Omega(\tau_{ij}-\tau^{f}_{i}) \pm \beta ) ]$, 
and the Laplace transforms of the impulse response functions $p_{i}(u)$ are given by $\hat{p}_{i}(\lambda) = \int_0^{\infty} \du p_{i}(u) e^{-\lambda u}= (1+\lambda b_{i})^{-a}$. 
\noindent
Rearranging and sorting for $d_1$ and $d_2$ yields 
\begin{equation}\label{eq:app-pert04}
d_1 =  
\left( \frac{ \alpha_{12} e^{-\lambda (\tau_{12} - \tau^{f}_{1})}}{\frac{\lambda}{\hat{p}_{1}(\lambda) e^{-\lambda \tau^{f}_{1}}} + \alpha_{12}} \right)
d_{2}, \;\;\;\;\; \text{and} \;\;\;\;\;\;
%
d_2 =  
\left( \frac{ \alpha_{21} e^{-\lambda (\tau_{21} - \tau^{f}_{2})}}{\frac{\lambda}{\hat{p}_{2}(\lambda) e^{-\lambda \tau^{f}_{2}}} + \alpha_{21}} \right)
d_{1}.
\end{equation}
\noindent
This expression can be written in matrix form as
\begin{equation} \label{eq:app-n=2-mchar1}
\zeta \cdot \mathbf{d} = \mathbf{G} \cdot  \mathbf{d}, 
\end{equation}
\noindent
where, $\zeta = 1$, vector $\mathbf{d} = (d_1, d_2)^{T}$ and the elements of the matrix $\mathbf{G}_{2 \times 2}$ are 
\begin{equation} \label{eq:app-n=2-jacobian2}
G_{kl} = 
\begin{cases}
& \left( \frac{ \alpha_{kl} e^{-\lambda (\tau_{kl} - \tau^{f}_{k})}}{\frac{\lambda}{\hat{p}_{k}(\lambda) e^{-\lambda \tau^{f}_{k}}} + \alpha_{kl}} \right),~\mbox{if}~ k \neq l, \vspace{1.0cm}\\
& 0,~~~~~~\mbox{if}~k = l; 
\end{cases}
\end{equation}
\noindent
for $k,l = 1,2$. From Eq.~\eqref{eq:app-n=2-mchar1}, we identify $\zeta$ $(= 1)$ as the eigenvalue of $\mathbf{G}$ which must satisfy the expression  
\begin{equation} \label{eq:app-n=2-mchar2}
\det{ \left( \mathbf{G} - \zeta \cdot \mathbf{I} \right) } = 0. 
\end{equation}
\noindent
From that we obtain the condition
\begin{equation} \label{eq:app-n=2-mchar3}
\begin{split}
1 - 
\left( \frac{ \alpha_{12} e^{-\lambda (\tau_{12} - \tau^{f}_{1})}}{\frac{\lambda}{\hat{p}_{1}(\lambda) e^{-\lambda \tau^{f}_{1}}} + \alpha_{12}} \right) 
\cdot
\left( \frac{ \alpha_{21} e^{-\lambda (\tau_{21} - \tau^{f}_{2})}}{\frac{\lambda}{\hat{p}_{2}(\lambda) e^{-\lambda \tau^{f}_{2}}} + \alpha_{21}} \right)
= 0, 
\end{split}
\end{equation}
which leads the following characteristic equation for the complex eigenvalues $\lambda=x+iy$
\vspace{0.1cm}
\begin{equation}\label{eq:app-n=2-stabcon2}
\begin{split}
\frac{\lambda^2}{\alpha_{12} \hat{p}_{1}(\lambda) \alpha_{21} \hat{p}_{2}(\lambda) e^{-2\lambda \bar{\tau}^{f}}} + 
\lambda 
\left( 
\frac{1}{ \alpha_{12} \hat{p}_{1}(\lambda) e^{-\lambda \tau^{f}_{1}} } + 
\frac{1}{ \alpha_{21} \hat{p}_{2}(\lambda) e^{-\lambda \tau^{f}_{2}} } 
\right)  
- \left( e^{-2\lambda (\bar{\tau} - \bar{\tau}^f)} - 1 \right)
=0.
\end{split}
\end{equation}
\vspace{0.1cm}
\noindent
From this equation we calculate the complex eigenvalues $\lambda = x + \rm{i}y$ and determine stability of the synchronized solutions characterized by collective frequency $\Omega$ and phase difference $\beta$, as discussed in section~\ref{sec:sync-sol-new}. 
\end{widetext}



\begin{widetext}

\part{Supplementary material}

\title{How heterogeneity affects synchronization and can enhance stability}
\author{Nirmal Punetha$^1$ and Lucas Wetzel$^{1,2}$}
\affiliation{$^1$Max-Planck-Institute for the Physics of Complex Systems, N\"othnitzer Stra{\ss}e 38, 01187 Dresden, Germany}
\affiliation{$^2$Center for Advancing Electronics Dresden, cfaed, W\"{u}rzburger Stra{\ss}e 46, 01187, Germany}

\maketitle 

\begin{center}
 \Large{Supplementary Material}
\end{center}

\section{Experiments and LTspice circuit level simulations} \label{supp-mat-exp-spice} 

\subsection{Experimental results from prototype setup}

The experimental results were obtained with a prototype setup of mutually delay-coupled digital phase-locked loops (DPLLs).
These DPLLs are made of \textit{CD4046B} integrated circuit-elements~\cite{CD4046B} with XOR phase detectors and a first-order RC loop-filter with a tunable \textit{Vishay Model 63} cermet trimmer resistor ranging from $R_{\rm min}=100$ Ohm to $R_{\rm max}=2$ MOhm with a tolerance of $\pm10\%$, see data-sheet~\cite{Vishay}.
The transmission delays are controlled using a \textit{Digilent ChipKit Max32} microcontroller~\cite{Digilent2011}.
Since generating significant transmission-delays in the milliseconds regime for the system of coupled PLLs with operational frequencies in the kilohertz regime would require connection lengths of the order of hundreds of kilometers, we use the microcontroller to buffer the signals of the PLLs and delay their output as specified.
The DPLL signals, phase-differences, transmission-delays and intrinsic and global frequencies were then measured using a \textit{PicoScope 2205} mixed-signal oscilloscope~\cite{Pico2014} and via the buffer history of the \textit{ChipKit Max32} microcontroller.
The same experimental setup had been used in~\cite{Wetzel2017}.

The experimental results presented in, e.g., Fig.~\ref{fig:new-exp-tau-fig11} and \ref{fig:new-exp-dtau-fig12} were obtained as follows.
In a first step we measured the transmission delay-times between the PLLs and obtained the mean and the difference for each bidirectional connection.
After transients had decayed, we captured $50$ digital waveforms of $10 \, \rm{ms}$, each with a sampling-rate of $1.22$MHz. 
The frequency $\Omega_j$ of the output signal can be calculated from the time-period $T_{{\Omega}_{j}}$, \ie the time-interval between two consecutive rising (or falling) edges of measurement $j$, where $j=1,\,\dots,\,J_T$ indexes all measurements in all $50$ waveforms that were recorded, see~Fig.~\ref{fig:supp-edges-freq-ph}. 
From these measurements we obtain sets of instantaneous $\Omega_j$ and $\beta_j$ values that are calculated from consecutive rising edges in each waveform for all $50$ waveforms. 
From this data-set, we calculate their respective mean values $\Omega$, $\beta$ and the corresponding errors. 
In order to calculate an average phase-difference $\beta$, we use the complex order parameter defined as $r_{\beta} \exp(\i \beta) = (J_T)^{-1} \sum_{j} \exp(\i \beta_j)$. 
As for large $J_T$ the value $r_{\beta}$ approaches zero if the angles $\beta_j$ are equally distributed between $[0,2\pi)$ and one if all $\beta_j$'s are equal, the value $(1-r_{\beta})$ was used to calculate the measurement error of $\beta$.
The error can then be specified in the range $[0,\,\pi)$ by multiplying the value $(1-r_{\beta})$ by $\pi$. 
Note that the incomplete cycles at the beginning and at the end of the signals and the transients after the system had been turned on were excluded.

\subsubsection{Heterogeneity in intrinsic frequencies and coupling strengths}

Using the results of the analysis of the phase model for mutually delay-coupled heterogeneous DPLL elements, we study the global frequencies and phase-differences as a function of the transmission delay.
The time-series of the signals were recorded and measured in a system of two mutually delay-coupled DPLL prototype units with detuned intrinsic frequencies and coupling strengths, see also table~\ref{tab:DPLLparam} in section~\ref{sec:exp-sim} of the paper.
From these measurements for different transmission-delays we obtained the times of rising and falling edges which we use to extract a phase. 
Here we made the assumption that the frequency is constant between two consecutive rising or falling edges and advance the phase linearly by $2\pi$ in the resulting time window, see Fig.~\ref{fig:supp-edges-freq-ph}.
Using this phase time-series we can then calculate the instantaneous frequencies of the DPLL elements, the global frequency of synchronized states, the Kuramoto order-parameter and the phase-differences.
The relation between the perturbation decay-rates, frequencies and the magnitude of the order-parameter $R(t)$ is given by
\begin{equation}
 \log\left[1-R(t)\right] = 2\sigma\,t+\log\left[ \cos^2(\gamma\,t) \right]+ C_0,
\end{equation}
where $\sigma=\textrm{Re}(\lambda)$, $\gamma=\textrm{Im}(\lambda)$ and $C_0$ a constant related to the initial perturbation vector~\cite{Wetzel2012}.
Hence, fitting the time-evolution of the Kuramoto order-parameter allows to measure the perturbation decay-rate and in the case of underdamped dynamics its associated frequency, which can then be compared to the theoretically predicted values.
However, due to the limitations of the current experimental setup the perturbation decay-rates could not measured here and will be addressed in the next-generation experimental setups that are planned.
The experimentally obtained results with the heterogeneous elements are in very good agreement with the values predicted by the phase model, see Fig.~\ref{fig:new-exp-tau-fig11} in  section~\ref{sec:exp-sim} of the paper.

\subsubsection{Heterogeneity in transmission-delays}

The transmission-delays are controlled using the microcontroller. 
Using the internal clock of the microcontroller the output signals are delayed by defined times and then sent out to the DPLLs in the network.
This can be done individually for each transmission-delay, i.e., for the signal transmission-delay from PLL $1$ to PLL $2$ and vice versa. 
For the experiments, the delays are changed such that the mean value of the transmission-delays remains constant.
For each value of delay-difference we measure the individual transmission-delays, record the signal time-series and the frequencies.
To confirm the value of transmission-delay set by the microcontroller we measured the time-difference between the same rising edge at the output of each PLL and the input at its coupling partner using the oscilloscope.
We plot the theoretical predicted values for the global frequencies and phase differences for DPLL elements with heterogeneous intrinsic frequencies and coupling strength.
The experimental results and corresponding LTspice simulations are in very good agreement with the results predicted by the phase model, see Figs.~\ref{fig:new-exp-tau-fig11}-\ref{fig:new-exp-dtau-fig12} in section~\ref{sec:exp-sim} of the paper.
\begin{figure}[h]
\includegraphics[scale=0.35, angle=0]{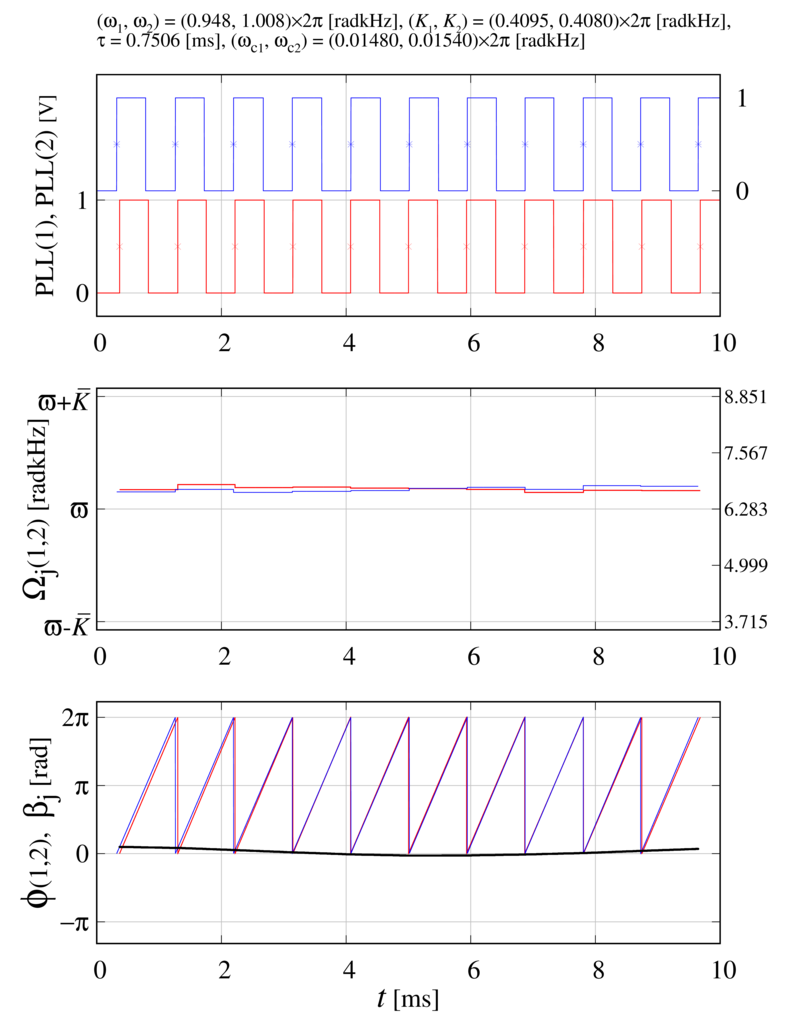}
 \caption{The digital waveforms of two DPLLs along with corresponding frequency and phase-difference.}
\label{fig:supp-edges-freq-ph}
\end{figure}

\subsubsection{Heterogeneity in cut-off frequencies}

The current experimental setup allows only a coarse-grained setting of the cut-off frequencies of the loop filter elements using tunable resistors.
These potentiometers range from $R_{\rm min}=100$ Ohm to $R_{\rm max}=2$ MOhm with a tolerance of $\pm10\%$ and can be tuned using a screwdriver, 
thereby changing the cut-off frequency of the loop filter.
The cut-off frequency is given by
\begin{equation}
 \omega_{}^{\rm c}=\frac{1}{R_{\rm LF} C_{\rm LF}},
\end{equation}
where the conductance of the RC loop-filter in our experimental setup is fixed at $C_{\rm LF}=22$nF and $R_{\rm LF}$ denotes the resistance.
We then measured signal time-series for the case of identical and different cut-off frequencies of the loop-filters, while we kept the mean cut-off frequency constant.
Transient measurements were also carried out, detuning the cut-off frequencies of the loop filters while the system was running.

In Fig.~\ref{fig:new-exp-dwc}, we show the results of these experiments, i.e., how heterogeneity in the cut-off frequencies of the loop-filters can be used to recover synchronization. 
For the case presented, the transmission delays $\tau_{12,21}$ were fixed at $0.7512 \, \rm{ms}$. 
We changed the cut-off frequencies $\omega_{1,2}^{c} = (R_{\rm LF} C_{\rm LF})^{-1}$ by adjusting resistance $R_{\rm LF}$ of the potentiometer. 
In this setup, we collected the digital waveforms of the two PLLs for $10 \, \rm{s}$ with a sampling-rate of $1$MHz and performed $20$ such measurements. 
Initially, the resistances were set to $R_{\rm LF}^1 = 131.2 \, \rm{k}\Omega$, $R_{\rm LF}^2 = 131.0 \, \rm{k}\Omega$, \ie loop-filters had approximately equal cut-off frequencies $(\omega_{1}^{\rm c},\omega_{2}^{\rm c})$ $=$ $(0.0551396, 0.0552238) \times 2\pi \, \rm{radkHz}$.  
At the beginning of our measurements, the DPLLs were desynchronized and the synchronized state was unstable. 
After approximately $4 \, \rm{s}$, we changed the resistances of the potentiometers to $R_{\rm LF}^1 = 490.0 \, \rm{k}\Omega$ and $R_{\rm LF}^2 = 75.6 \, \rm{k}\Omega$ such that the the new cut-off frequencies were $\omega_{1}^{\rm c}=0.0148 \times 2\pi \, \rm{radkHz}$ and $\omega_{2}^{\rm c}=0.0957 \times 2\pi \, \rm{radkHz}$ respectively. 
This change increased the heterogeneity in the cut-off frequencies while the mean value remained approximately the same ($\bar{\omega}_{c} = 0.05524 \times 2\pi$). 
As a result, the system synchronized to a state with a common global frequency, see Fig.~\ref{fig:new-exp-dwc} in section~\ref{sec:exp-sim} of the paper.

\subsection{LTspice simulations}

We validate the theoretical results obtained from the phase model using LTspice simulations~\cite{LTspice}. 
This uses circuit level models for the dynamics of voltages, currents and processing times of standard electronic components, most of which are available on the market~\cite{LTspice}.
The operational amplifiers used in the adders are \textit{ADA4857} components~\cite{ADA4857}.
As a VCO we use the \textit{LTC6900} oscillator~\cite{LTC6900}.
For this device we obtain the input sensitivity and measure the operation frequency as a function of the input voltage $V_{\rm bias}$, see Fig.~\ref{fig:LTC6900}.
From this we then obtained the input sensitivity of the VCO as $K^{\rm VCO}=0.088291$MHz/V.
\begin{figure}[b]
\includegraphics[scale=0.158, angle=0]{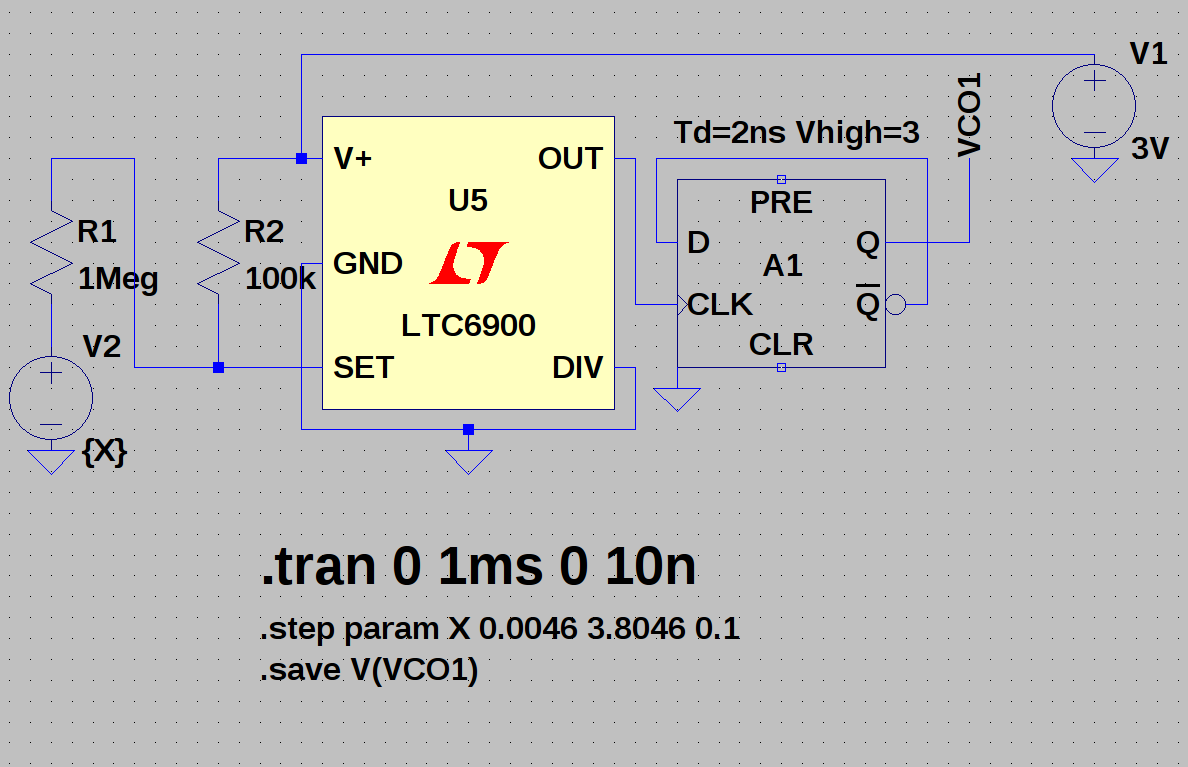} \hspace{1cm}
\includegraphics[scale=0.28, angle=0]{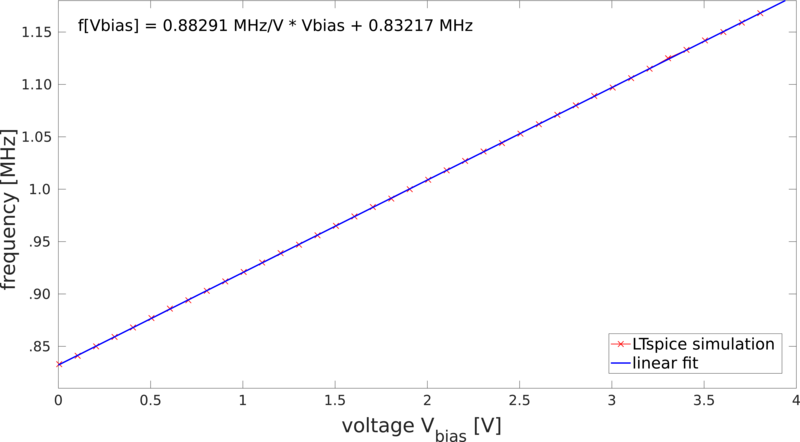}
 \caption{SPICE model circuitry of LTC6900 oscillator with supply voltage to measure the input sensitivity. }
\label{fig:LTC6900}
\end{figure}
\begin{figure}[ht]
\includegraphics [scale=0.28, angle=0]{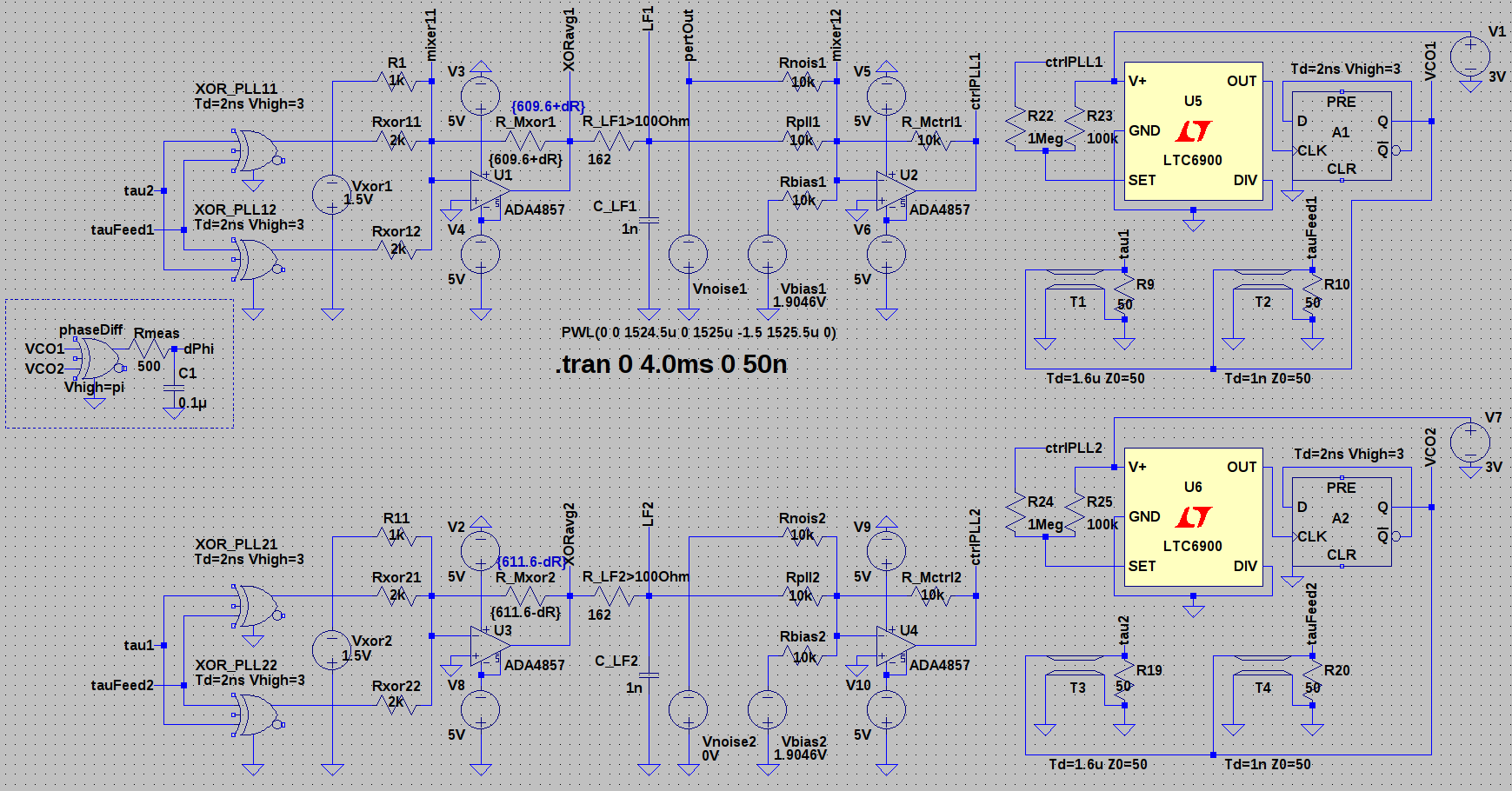}
 \caption{SPICE model circuitry in LTspice for simulations.}
\label{fig:sketchLTspiceCircuit}
\end{figure}
The frequency of the VCO depends of the offset voltage $V_{\rm bias}$ as
\begin{equation}
 f(V_{\rm bias}) \left[ \textrm{MHz} \right] = 0.088291\, \left[\textrm{MHz/V}\right]\times V_{\rm bias}\left[ \textrm{V} \right] +0.83217\, \left[\textrm{MHz}\right].
\end{equation}

Then we use a setup of two mutually delay-coupled DPLLs to simulate self-organized synchronization on the circuit level, see Fig.~\ref{fig:sketchLTspiceCircuit}.
We measure and monitor the time-series of the VCO's output voltages, the control signals, the loop-filter signals and the perturbation signals. 
The perturbation signals are added to the control signal of one of the DPLLs and cause phase-deviations from the synchronized state.
Using the time-series of the output-signals of the VCOs we can obtain the phases, phase-differences and the instantaneous frequencies.
From these quantities the Kuramoto order-parameter can be calculated as a measure of the synchrony in the system.
Also, the perturbation decay rates can be measured from the time-series of the phase-differences or the time-evolution of the order parameter~\cite{Wetzel2012}.
A network of two DPLLs is characterized by the intrinsic frequencies $\omega_k$ and the coupling strengths $K_k$ of the VCOs, the transmission- and feedback-delays $\tau_{kl},\,\tau_{kl}^f$, and the cut-off frequencies of the loop filters $\omega_{k}^{\rm c}$.
In the LTspice model, these parameters are set by different electronic components.
The cut-off frequencies are set using the resistors $R_{\textrm{LF},\,k}$ of the RC-circuit.
The coupling strengths are set by gains in the adder of the XOR signals ($S_{k1}$) and the adder for the bias voltage to the loop-filter signal ($S_{k2}$).
These gains $S_k$ are set by the resistors $R_{\textrm{MXOR},\,k}$ and $R_{\textrm{PLL},\,k}$, where we use the first gain to set the coupling strength and the second gain compensates for damping in the low-pass filter.
We derived the corresponding relations of the system parameters and the LTspice model and found for the total and individual gains
\begin{eqnarray}
 S_{k1} &=& S_k / S_{k2}, 				\\
 S_{k2} &=& \frac{R_{\textrm{PLL},\,k}\left[ \textrm{Ohm} \right]}{R_{\textrm{PLL},\,k}\left[ \textrm{Ohm} \right]+R_{\textrm{LF},\,k}\left[ \textrm{Ohm} \right]}, \\
 S_{k}  &=& S_{k1}S_{k2} = \frac{2K \left[ \textrm{radHz} \right]}{A\left[ \textrm{V} \right]\times K^{\rm VCO}\left[ \textrm{radHz/V} \right]},	
\end{eqnarray}
where $A$ denotes the output amplitude $V_{\rm high}$ of the XORs, and $k=1,2$ indexes the DPLLs.
The resistance-values of the resistors of the loop-filter cut-off frequency $\omega_k^{\rm c}$ and gain $S_{k1}$ of the phase-detector signal can be calculated from
\begin{eqnarray}
 R_{\textrm{MXOR},\,k} &=& \frac{K \left[ \textrm{radHz} \right]\times R_{\textrm{XOR},\,k}\left[ \textrm{Ohm} \right] }{A\left[ \textrm{V} \right]\times K^{\rm VCO}\left[ \textrm{radHz/V} \right]\times S_{k2}}\\
 R_{\textrm{LF},\,k}   &=& \frac{R_{\textrm{PLL},\,k}\left[ \textrm{Ohm} \right]}{-1+\omega_k^{\rm c}\left[ \textrm{radHz} \right]\times R_{\textrm{PLL},\,k}\left[ \textrm{Ohm} \right]\times C_{\rm LF}\left[ \textrm{F} \right]},
\end{eqnarray}
where $C_{\rm LF}$ is the capacitance of the loop-filter capacitor.

\section{Numerical simulation of the phase model} \label{supp:simulation}

\subsection{An $a$th order filter: PLL dynamics as $(a+1)$th order differential equation} \label{supp-mat-second-order-km}

\noindent
In the following, we show that system dynamics with $a$th order filter can be 
written as $(a+1)$th-order differential equation. For this, consider the 
dynamics of $N$ delay-coupled PLLs, 
\begin{equation} \label{eq:sys0}
\dot{\phi}_k = \omega_k + K_{k} \int_0^{\infty} \du p(u;a,b) ~\xkpd(t-u);~~k=1,\cdots N.
\end{equation}
\noindent
rewriting this equation as 
\begin{equation} \label{eq:sys00}
\begin{split}
\dot{\phi}_k = \omega_k + K_{k} ~\left(p * \xkpd \right) (t);~~k=1,\cdots N, 
\end{split}
\end{equation}
\noindent
where, ($p * \xkpd$) represents the convolution operation on $p$ and $\xkpd$. 
Taking the Laplace transform ($\mathcal{L}$) of this equation, and using the property 
$\mathcal{L}\left[(f*g)(t)\right] = \mathcal{L}(f(t))\cdot\mathcal{L}(g(t))$, 
we get,
\begin{equation} \label{eq:laplace_eq01}
\begin{split}
\mathcal{L} ( \dot{\phi}_k(t) ) =& ~\mathcal{L}(\omega_k) 
+ K_{k} ~\mathcal{L}(p(t)) \cdot \mathcal{L}(\xkpd (t)),\\
\Rightarrow~ \hat{\dot{\phi}}_k(s) =& ~\frac{\omega_k}{s} 
+ K_{k} ~\hat{p}(s) \cdot \hat{x}_k^{\text{PD}}(s).
\end{split}
\end{equation} 
\noindent
Here hats ( $\hat{•}$ ) represent the Laplace transforms of the respective functions. 
Since $\hat{\dot{\phi}}_k(s) = s {\hat{\phi}}_k (s) - {\phi}_k (0)$, one can rewrite this 
equation as, 
\begin{equation} \label{eq:laplace_eq02}
s {\hat{\phi}}_k (s) - {\phi}_k (0) = \frac{\omega_k}{s} 
+ K_{k} ~\hat{p}(s) \cdot \hat{x}_k^{\text{PD}}(s).
\end{equation} 
\noindent
Laplace transform of the impulse response function $\hat{p}(s)$ for $a$th order filter is given by, 
\begin{equation}\label{eq:pu03}
\hat{p}(s) = \frac{1}{(s b + 1)^a}. 
\end{equation} 
\noindent
Substituting this value in Eq.~\eqref{eq:laplace_eq02}, we obtain, 
\begin{equation} \label{eq:laplace_eq03}
\begin{split}
s {\hat{\phi}}_k (s) - {\phi}_k (0) = \frac{\omega_k}{s} + K_{k} ~\frac{1}{(s b + 1)^a} \cdot \hat{x}_k^{\text{PD}}(s).
\end{split}
\end{equation} 
\noindent
Therefore, with $a$th order filter, dynamics of the system in Laplace space is given by 
\begin{equation} \label{eq:laplace_eq03a}
\begin{split}
(s b + 1)^a \left[ s {\hat{\phi}}_k (s) - {\phi}_k (0) \right] = 
\frac{\omega_k}{s} (s b + 1)^a + K_{k}\cdot \hat{x}_k^{\text{PD}}(s).
\end{split}
\end{equation} 
\noindent
Substituting binomial expansion for $(s b + 1)^a$ \ie 
\begin{equation} \label{eq:binomial-exp}
\begin{split}
(sb + 1)^a = \sum_{r=0}^{a} ~\Comb{a}{r}~ (sb)^r,\\
\mbox{where,}~~~ \Comb{a}{r} = \Comb{a}{a-r} = \frac{a!}{r!~(a-r)!},
\end{split}
\end{equation}
\noindent
we get,
\begin{equation} \label{eq:laplace_eq03b}
\begin{split}
\sum_{r=0}^{a} ~\Comb{a}{r}~ b^r s^r \left[ s {\hat{\phi}}_k (s) - {\phi}_k (0) \right] = \frac{\omega_k}{s} \sum_{r=0}^{a} ~\Comb{a}{r}~ b^r s^r + K_{k} \cdot \hat{x}_k^{\text{PD}}(s),
\end{split}
\end{equation}
\noindent
\begin{equation} \label{eq:laplace_eq03c}
\begin{split}
\Rightarrow \sum_{r=0}^{a} ~\Comb{a}{r}~ b^r s^{r+1} \hat{\phi}_k (s) - \sum_{r=0}^{a} ~\Comb{a}{r}~ b^r s^r {\phi}_k (0) = \frac{\omega_k}{s} \sum_{r=0}^{a} ~\Comb{a}{r}~ b^r s^r + K_{k} \cdot \hat{x}_k^{\text{PD}}(s).
\end{split}
\end{equation}
\noindent
Using the property that the Laplace transform of $n$th-order derivative of a function (say $f^{[n]}(t)$) is   
\begin{equation} \label{eq:laplace-deri}
\mathcal{L}[f^{[n]}(t)] = \hat{f}^{[n]}(s) = s^n \hat{f}(s) - \sum_{j=1}^{n} s^{n-j} f^{[j-1]}(0), 
\end{equation}
\noindent
Eq.~\eqref{eq:laplace_eq03c} reads, 
\begin{equation} \label{eq:laplace_eq03d}
\begin{split}
\sum_{r=0}^{a}& ~\Comb{a}{r}~ b^r 
\left( \hat{\phi}^{[r+1]}_{k}(s) + \sum_{j=1}^{r+1} s^{r+1-j} {\phi}^{[j-1]}_{k}(0) \right)
- \sum_{r=0}^{a} ~\Comb{a}{r}~ b^r s^r {\phi}_k (0) = 
\frac{\omega_k}{s} + \omega_k \sum_{r=1}^{a} ~\Comb{a}{r}~ b^r s^{r-1} 
+ K_{k} \cdot \hat{x}_k^{\text{PD}}(s),\\
\Rightarrow \sum_{r=0}^{a}& ~\Comb{a}{r}~ b^r \hat{\phi}^{[r+1]}_{k}(s) + \left[ {\phi}_{k}(0) + 
\sum_{r=1}^{a} ~\Comb{a}{r}~ b^r \left( \sum_{j=1}^{r+1} s^{r+1-j} {\phi}^{[j-1]}_{k}(0) \right) \right] 
- \left[ {\phi}_{k}(0) + \sum_{r=1}^{a} ~\Comb{a}{r}~ b^r s^r {\phi}_k (0) \right]\\
&~~~~~~~~~~~~~~~~~~~~~~~~~~~~~~~~~~~~~~~~~~~~~~~~~~~~~~~~~~~~~~~~= 
\frac{\omega_k}{s} + \omega_k \sum_{r=1}^{a} ~\Comb{a}{r}~ b^r s^{r-1} 
+ K_{k} \cdot \hat{x}_k^{\text{PD}}(s),\\
\Rightarrow \sum_{r=0}^{a}& ~\Comb{a}{r}~ b^r \hat{\phi}^{[r+1]}_{k}(s) = 
\frac{\omega_k}{s} 
+\sum_{r=1}^{a} ~\Comb{a}{r}~ b^r \left( -\sum_{j=1}^{r+1} s^{r+1-j} {\phi}^{[j-1]}_{k}(0)
 +  s^r {\phi}_k (0)
 + s^{r-1} \omega_k \right)
+ K_{k} \cdot \hat{x}_k^{\text{PD}}(s).
\end{split}
\end{equation}
\noindent
Inverse Laplace transform of above equation back into the time domain gives, 
\begin{equation} \label{eq:laplace_eq03f}
\begin{split}
\sum_{r=0}^{a} ~\Comb{a}{r}~ b^r 
\frac{ \mathrm{d}^{r+1} \phi_{k}(t) }{ \mathrm{d} t^{r+1} } = 
\omega_k + P_k^{\delta}(a)  + K_{k} \cdot x_k^{\text{PD}}(t),\\
\end{split}
\end{equation}
\noindent
where, 
\begin{equation} \label{eq:laplace_eq03g}
\begin{split}
P_k^{\delta}(a) = -\sum_{r=1}^{a} ~\Comb{a}{r}~ b^r 
\left( \sum_{j=1}^{r+1} {\phi}^{[j-1]}_{k}(0) \cdot \delta^{[r+1-j]}(t) \right)
+ \sum_{r=1}^{a} ~\Comb{a}{r}~ b^r \left[ {\phi}_k (0) \cdot \delta^{[r]}(t)
+ \omega_k \cdot \delta^{[r-1]}(t)
 \right].
\end{split}
\end{equation}
\noindent
Here $\delta$(t) represents the Dirac-delta function and $\delta^{[n]}(t)$ its $n$th 
derivative. Hence, the dynamics of the system with $a$th order filter is 
governed by $(a+1)$th order differential equation, which is given by Eq.~\eqref{eq:laplace_eq03f}. 
Few examples ($a=0,1,2,3$) are given below. \\ 
\noindent
{\bf Filter of order zero ($a=0$): } 
For a zero-th order filter, $P_k^{\delta}(0) = 0$, and the frequencies of the VCOs are 
given by a simple first order differential equation,
\begin{equation} \label{eq:a=0a}
\begin{split}
\frac{ \mathrm{d}\phi_{k}(t) }{ \mathrm{d}t} = \omega_k + K_{k} \cdot x_k^{\text{PD}}(t).\\
\end{split}
\end{equation}
\noindent

\noindent
{\bf Filter of order one ($a=1$): } 
Now for a first order filter \ie $a=1$, 
\begin{equation} \label{eq:a=1a}
\begin{split}
P_k^{\delta}(1) =&~ -b \left( {\phi}_{k}(0) \cdot \delta^{[1]}(t) + {\phi}^{[1]}_{k}(0) \delta (t) \right)
+ b \left[ {\phi}_{k}(0) \cdot \delta^{[1]}(t) + \omega_k \cdot \delta {(t)} \right]\\
=&~ -b \left( {\phi}^{[1]}_{k}(0) - \omega_k \right) \cdot \delta {(t)}. 
\end{split}
\end{equation}
\noindent
Therefore, the phase dynamics for the first order filter is given by
\begin{equation} \label{eq:a=1b}
\begin{split}
b~\frac{ \mathrm{d}^{2} \phi_{k}(t) }{ \mathrm{d} t^{2} } + \frac{ \mathrm{d}\phi_{k}(t) }{ \mathrm{d}t} = 
\omega_k + b \left( \omega_k - {\phi}^{[1]}_{k}(0) \right) \cdot \delta {(t)}  + K_{k} \cdot x_k^{\text{PD}}(t).\\
\end{split}
\end{equation}
\noindent

\noindent
{\bf Filter of order two ($a=2$): } For second order filter
\begin{equation} \label{eq:a=2a}
\begin{split}
P_k^{\delta}(2) =&~ -2 b \left( {\phi}_{k}(0) \cdot \delta^{[1]}(t) + 
{\phi}^{[1]}_{k}(0) \cdot \delta{(t)} \right) 
- b^2 \left( {\phi}_{k}(0) \cdot \delta^{[2]}(t) + 
{\phi}^{[1]}_{k}(0) \cdot \delta^{[1]}(t) + 
{\phi}^{[2]}_{k}(0) \cdot \delta{(t)} \right) \\
&~~~~~+ 2 b \left[ {\phi}_{k}(0) \cdot \delta^{[1]}(t) + \omega_k \cdot \delta{(t)} \right]
+ b^2 \left[ {\phi}_{k}(0) \cdot \delta^{[2]}(t) + \omega_k \cdot \delta^{[1]}(t) \right],\\
\\
=&~ -2 b~ {\phi}^{[1]}_{k}(0) \cdot \delta{(t)}
- b^2 \left( 
{\phi}^{[1]}_{k}(0) \cdot \delta^{[1]}(t) + 
{\phi}^{[2]}_{k}(0) \cdot \delta{(t)} \right) + 2 b~ \omega_k \cdot \delta{(t)} + 
b^2~ \omega_k \cdot \delta^{[1]}(t),\\
\\
=&~ 2 b \left( \omega_k - {\phi}^{[1]}_{k}(0) \right) \cdot \delta{(t)} + 
b^2 \left( \omega_k - {\phi}^{[1]}_{k}(0) \right) \cdot \delta^{[1]}(t) - 
b^2 {\phi}^{[2]}_{k}(0) \cdot \delta{(t)}.
\end{split}
\end{equation}
\noindent
The dynamics is govern by the third order differential equation
\begin{equation} \label{eq:a=2b}
\begin{split}
b^2 \frac{ \mathrm{d}^{3} \phi_{k}(t) }{ \mathrm{d} t^{3} } + 2b \frac{ \mathrm{d}^{2} \phi_{k}(t) }{ \mathrm{d} t^{2} } + \frac{ \mathrm{d}\phi_{k}(t) }{ \mathrm{d}t} = 
\omega_k + P_k^{\delta}(2) + K_{k} \cdot x_k^{\text{PD}}(t).\\
\end{split}
\end{equation}
Therefore, depending on the order of the loop-filter $a$, one can rewrite the integro delay-differential equation for the phase-dynamics of coupled PLLs system as $(a+1)$th order delay differential equation.

\begin{figure}
\hspace{0.1cm}(a)\hspace{8cm}(b)\\
\includegraphics [scale=0.30, angle=0]{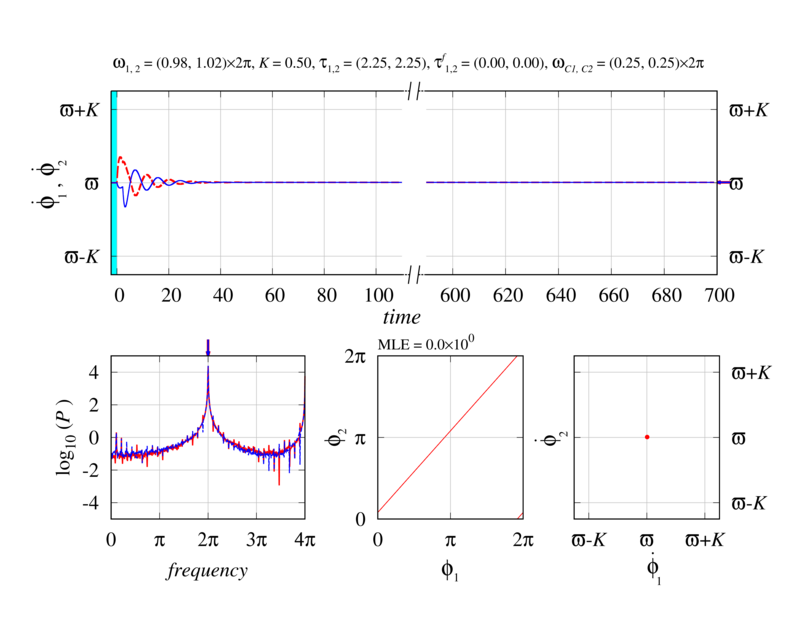}
\includegraphics [scale=0.30, angle=0]{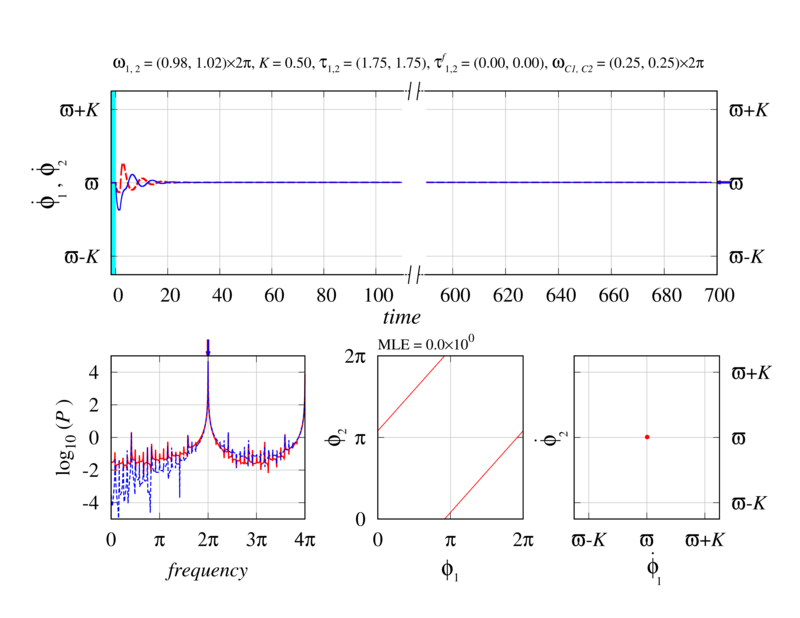}\\
\hspace{0.1cm}(c)\hspace{8cm}(d)\\
\includegraphics [scale=0.30, angle=0]{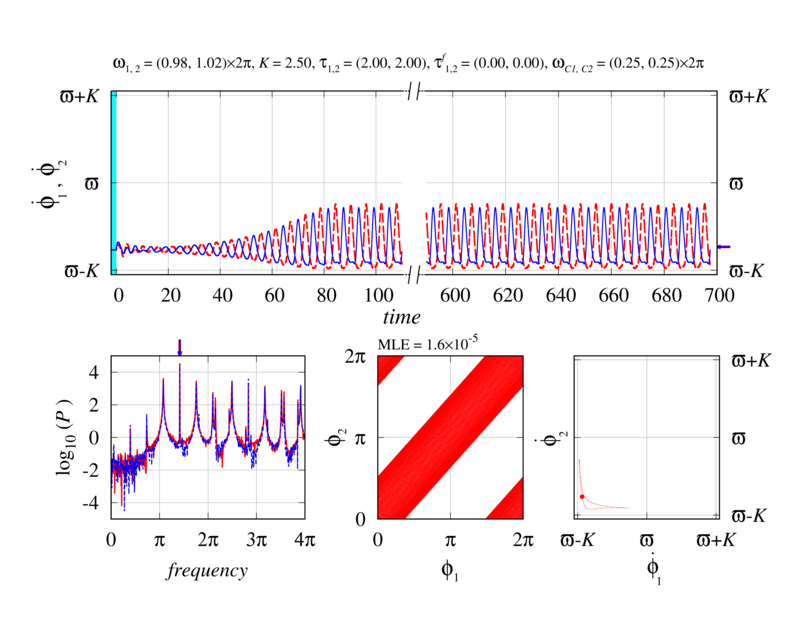}
\includegraphics [scale=0.30, angle=0]{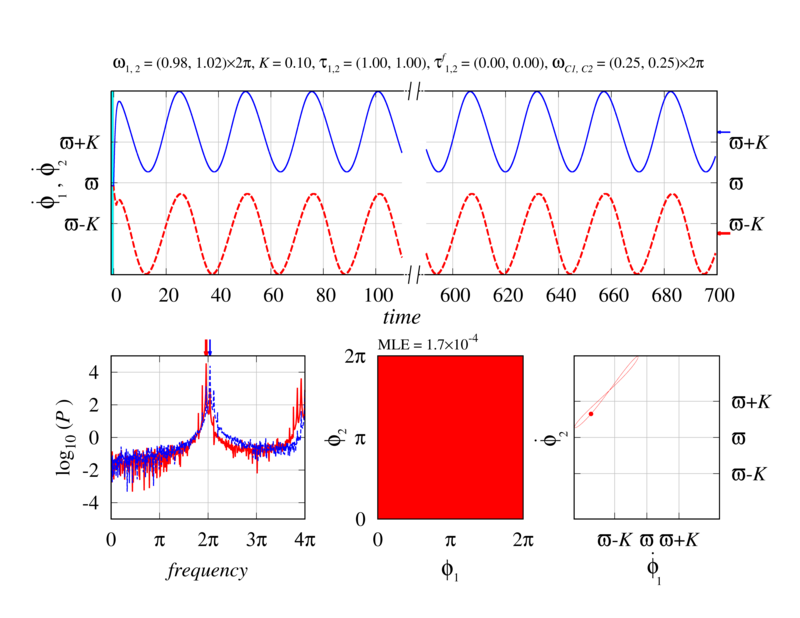}
 \caption{(Color online) 
 Simulation results obtained from the second order differential equation Eq.~\eqref{eq:reduct-pd3} for the coupled PLLs with first order filters. 
 Time variation of the frequencies $\dot{\phi}_{1,2}$ of two delay-coupled PLL clocks and shown at top subfigure. 
 Fourier spectrum of the time-series, the phase ($\phi_1-\phi_2$ plane) and the frequency ($\dot{\phi}_1-\dot{\phi}_2$) relations are shown at the bottom left, bottom middle and bottom right column respectively. 
 The parameter values are mentioned at the top of the figure. 
 For the parameter values given in (a) and (b), the system has stable modified-inphase and stable modified-antiphase synchronized solution respectively. 
 If synchronized solutions exist but are unstable (c) or do not exist at all (d), the oscillators have time dependent frequencies and the system is desynchronized. 
}
\label{fig:supp-ht-plot-fft}
\end{figure}
\begin{figure}
\includegraphics [scale=0.5, angle=0]{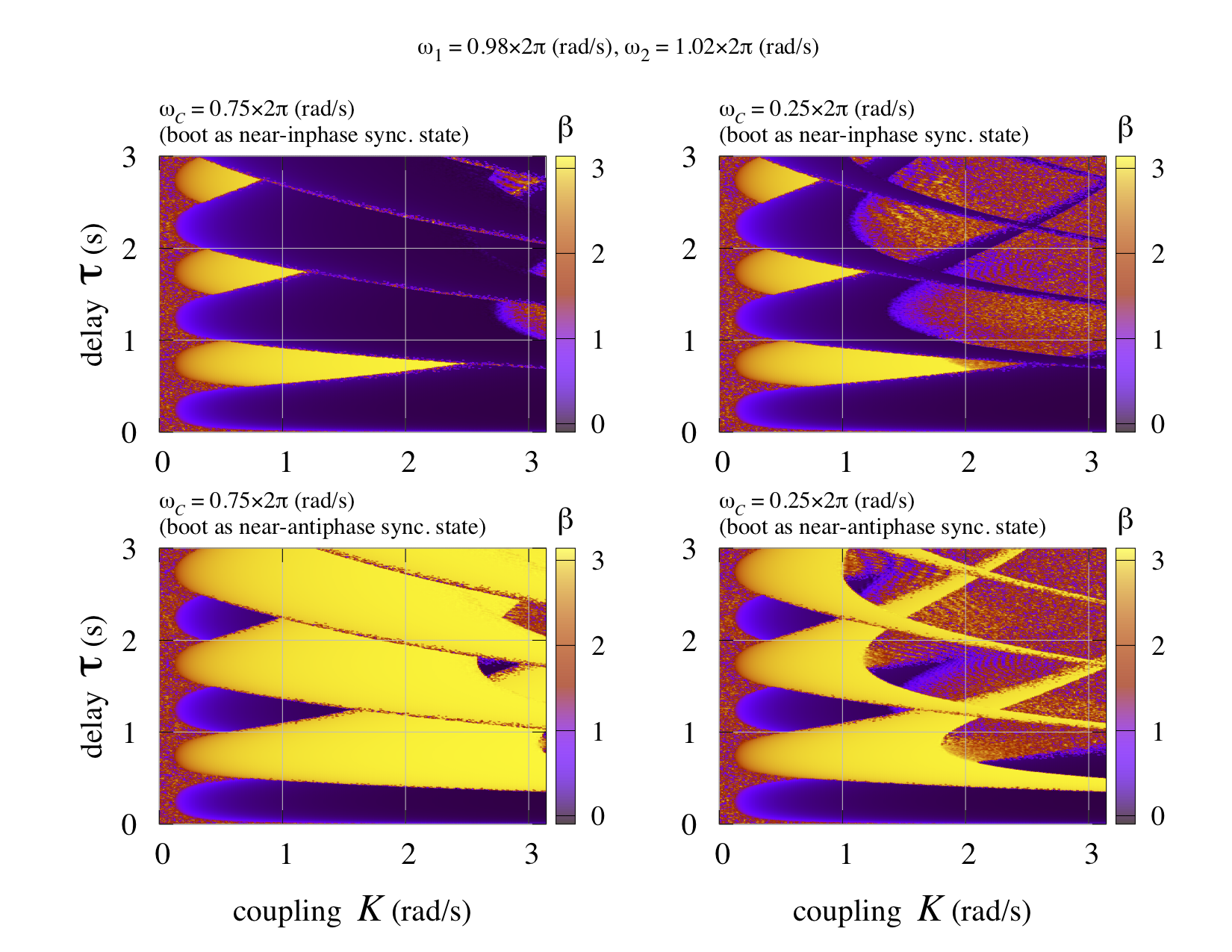}
 \caption{(Color online) 
Phase difference $\beta$ from the simulation plotted in the parameter ($K, \tau$) space for two $N=2$ delay-coupled PLLs at two different cut-off frequencies $\omega_c = 0.75 \times 2\pi$ (left column) and $\omega_c = 0.25 \times 2\pi$ (right column). 
The intrinsic frequencies are $\omega_{1,2} = (1 \mp 0.02) \times 2\pi$. 
We use near in-phase (upper) and near antiphase (lower) solutions as initial conditions. 
We simulate the system with small perturbation in the phase $\delta[\phi_{ini}] = 2\pi/100$ at the time $t=0$, and determine the final state of the system numerically. 
}
\label{fig:supp-ktau-plain}
\end{figure}

\subsection{Simulation results with second order differential equation} 

For our study we consider the PLLs with first order loop filter. 
For this system, the phase dynamics can be written as 
\begin{equation} \label{eq:reduct-pd1}
\begin{split}
b~\frac{ \mathrm{d}^{2} \phi_{k}(t) }{ \mathrm{d} t^{2} } + \frac{ \mathrm{d}\phi_{k}(t) }{ \mathrm{d}t} = 
\omega_k + b \left( \omega_k - {\phi}^{[1]}_{k}(0) \right) \cdot \delta {(t)}  + K_{k} \cdot x_k^{\text{PD}}(t),\\
\end{split}
\end{equation}
\noindent
see Eq.~\eqref{eq:a=1a}.
The phase detector signals $x_k^{\text{PD}}(t)$ for analog and digital PLLs are, 
\begin{eqnarray} \label{eq:reduct-pd2}
 x_k^{\rm PD,analog}  &=&  \frac{1}{n_k}\sum\limits d_{kl} \cos\left( \phi_l(t-\tau_{kl}) - \phi_k(t-\tau_{kl}^f) \right), \\
 x_k^{\rm PD,digital} &=&  \frac{1}{2}+\frac{1}{2n_k}\sum\limits d_{kl} \Delta\left( \phi_l(t-\tau_{kl}) - \phi_k(t-\tau_{kl}^f) \right).
\end{eqnarray}
Therefore, the dynamics of delay-coupled analog PLLs system is given by 
\begin{equation} \label{eq:reduct-pd3}
\begin{split}
b~\ddot{\phi}_{k} + \dot{\phi}_{k} = \omega_k + b \left( \omega_k - {\phi}^{[1]}_{k}(0) \right) \cdot \delta {(t)} + \frac{K_k}{n_k}\sum\limits d_{kl} \cos\left( \phi_l(t-\tau_{kl}) - \phi_k(t-\tau_{kl}^f) \right).\\
\end{split}
\end{equation}
\noindent
We use this second order differential equation for our simulations and examine different dynamical scenarios exhibited by the system at different parameter values. 
One can choose the parameter values and the history for initial conditions and see how the system settles into a modified-inphase, modified-anti-phase states, or remains desynchronized (either synchronized states are unstable or it do not exist). 
Examples of these cases are shown in the Supplementary Figs.~\ref{fig:supp-ht-plot-fft} (a)(b)(c) and (d) respectively. 
These simulations are helpful in quickly verifying the analytical results and to get insight into the behavior of the system. 
The results shown in Supplementary Fig.~\ref{fig:new-hf-fig14} indicate the possibility of different dynamical scenarios in parameter space. 
Here, initial conditions play important role due to the multistability. 
This behavior can be examined numerically as in Supplementary Fig.~\ref{fig:supp-ktau-plain}, where we show the simulation results for the asymptotic phase dynamics in the ($K,\tau$) plane. 
In the multistable regions where both modified in-- and antiphase solutions coexist, simulations show how the final state of the system depends on the initial conditions. 
As multiple stable synchronized states may coexist in the system, simulations are also helpful to examine the basin of attraction for these multiple solutions. 
%

\section{Extension to the network of $N$ delay-coupled oscillators} \label{supp-mat-Nosc}
Sec.~\ref{sec:sync-sol-new} in the paper provides the analysis for two delay-coupled oscillators with heterogeneous components, where we obtain the expressions to evaluate synchronized solutions and their stability. 
These transcendental equations can be solved numerically to calculate the global frequency, the phase difference and the corresponding perturbation response rates. 
For a larger network of delay-coupled PLLs ($N>2$) with heterogeneous components, the analysis can be extended as presented in the following. 

\subsection{Synchronized solutions}

\noindent
The dynamics of $N$ delay-coupled PLLs, in a general form, is given by, 
\begin{equation} \label{eq:supp-sys-gencpl}
\dot{\phi}_k(t) = \omega_k + \frac{K_k}{n_k} \sum_{l=1}^N c_{kl} 
\int_0^{\infty} \du p_{k}(u)~ h \left[ \phi_l (t-u-\tau_{kl}) - \phi_k (t - u -\tau^{f}_{k}) \right].
\end{equation} 
\noindent
Synchronized solutions are given by the ansatz 
\begin{equation}\label{eq:supp-sync-ansatz}
\phi_k = \Omega t + \beta_k, ~~k = 1,2, \cdots,N,
\end{equation}
\noindent
where $\Omega$ is the common collective frequency and $\beta_k$ is the phase deviation. 
Putting these solutions in Eq.~\eqref{eq:supp-sys-gencpl}, we get 
\begin{equation}\label{eq:supp-sync-gen-con}
\frac{\text{d} \phi_k}{\text{d}t} = \Omega = \omega_k + \frac{K_k}{n_k}\sum_{l=1}^{N} 
c_{kl}~h \left[ -\Omega (\tau_{kl} - \tau^f_{k}) - \beta_{kl} \right], ~~~~~k=1,\cdots,N,
\end{equation}
\noindent 
were $ \beta_{kl}=(\beta_{k} - \beta_{l})$. 
Eqs.~\eqref{eq:supp-sync-gen-con} provide condition for the synchronized solutions described in Eq.~\eqref{eq:supp-sync-ansatz}. 
Note that, there are $N$ equations, see Eqs.~\eqref{eq:supp-sync-gen-con}, to be solved simultaneously for ($N+1$) values, $\lbrace \Omega, \beta_1, \beta_2 \cdots, \beta_N \rbrace$. 
This can be reduced to $N$ equations simply be assuming $\beta_1 = 0$ as a reference. 
The $\beta$-values then represent the phase differences of individual oscillators with respect to the reference oscillator. 
These $N$ values (\ie the solutions $\lbrace \Omega, \beta_2 \cdots, \beta_N \rbrace$) can then be calculated from the following $N$ number of equations. 
\begin{equation}\label{eq:supp-sync-gen-con1}
\begin{split}
\Omega =&~ \omega_1 + \frac{K_1}{n_1}\sum_{l=1}^{N} 
c_{1l}~h \left[ -\Omega (\tau_{1l} - \tau^{f}_{1}) - \beta_{1l} \right],\\ 
=&~ \omega_2 + \frac{K_2}{n_2}\sum_{l=1}^{N} 
c_{2l}~h \left[ -\Omega (\tau_{2l} - \tau^{f}_{2}) - \beta_{2l} \right],\\ 
\vdots\\
=&~ \omega_N + \frac{K_N}{n_N}\sum_{l=1}^{N} 
c_{Nl}~h \left[ -\Omega (\tau_{Nl} - \tau^{f}_{N}) - \beta_{Nl} \right]. 
\end{split}
\end{equation}
\noindent
Therefore, the phase-locked solutions satisfy, 
\begin{equation}\label{eq:supp-sync-gen-con2}
\begin{split}
\omega_1 + \frac{K_1}{n_1} S_1 = 
\omega_2 + \frac{K_2}{n_2} S_2 = \hdots =
\omega_N + \frac{K_N}{n_N} S_N;\\
\mbox{where,}~~ S_k = \sum_{l=1}^{N} 
c_{kl}~h \left[ -\Omega (\tau_{kl} - \tau^{f}_{k}) - \beta_{kl} \right]. 
\end{split}
\end{equation}
\noindent
A compact form of the condition for synchronized solutions can be obtained by averaging over all oscillators, 
\begin{equation}\label{eq:supp-sync-gen-avg}
\begin{split}
\Omega = \bar{\omega} + 
\frac{1}{N} \sum_{k=1}^{N} \left( K_k \sum_{l=1}^{N} 
\frac{c_{kl}}{n_k}~h \left[ -\Omega (\tau_{kl} - \tau^{f}_{k}) - \beta_{kl} \right] \right),
\end{split}
\end{equation}
\noindent 
These equations provide synchronized solutions of a notwork of delay-coupled oscillators for a given coupling function and connection topology. 
For larger $N>2$ it is nontrivial to obtain explicit analytic expressions for synchronized frequency and phase differences. 
One can use, however, for example, linear approximations for coupling function or numerical methods to estimate these solutions.

\subsection{Stability of the synchronized solutions} \label{supp:model-N-osci}

For a system of $N$ oscillators, we here obtain a general expression for the stability of synchronized solutions. 
In order to calculate linear stability, we apply small perturbations $q_{k}$ to the synchronized solutions, \ie 
\begin{equation}\label{eq:supp-small-pert}
\phi_k = \Omega t + \beta_{k} + q_k; ~~k = 1, 2, \cdots N;  
\end{equation}
\noindent
and examine the behavior of these perturbations. Putting these values in Eq.~\eqref{eq:supp-sys-gencpl}, we obtain, 
\begin{equation}\label{eq:supp-pert00}
\begin{split}
\Omega + \dot{q}_{k} = \omega_{k} + \frac{K_{k}}{n_k} \sum_{l=1}^{N} c_{kl} \int_0^{\infty} \du p_{k}(u) ~h \left[ \left( -\Omega(\tau_{kl}-\tau^{f}_{k}) - \beta_{kl} \right) 
+ \left( q_{l}^{u+\tau_{kl}} - q_{k}^{u+\tau^{f}_{k}} \right) \right].
\end{split}
\end{equation}
\noindent
Expanding function `$h$' to the first order, we have 
\begin{equation}\label{eq:supp-pert01}
\begin{split}
\Omega + \dot{q}_{k} = \omega_{k} +&~ \frac{K_{k}}{n_k} \sum_{l=1}^{N} c_{kl} \int_0^{\infty} \du p_{k}(u) 
~h \left[ \left( -\Omega(\tau_{kl}-\tau^{f}_{k}) - \beta_{kl} \right) \right] \\
+ \frac{K_{k}}{n_k}& \sum_{l=1}^{N} c_{kl} \int_0^{\infty} \du p_{k}(u) ~h^{\prime} \left[ \left( -\Omega(\tau_{kl}-\tau^{f}_{k}) - \beta_{kl} \right) \right] \left( q_{l}^{u+\tau_{kl}} - q_{k}^{u+\tau^f_{k}} \right). 
\end{split}
\end{equation}
\noindent
Therefore, the perturbation dynamics is given by 
\begin{equation}\label{eq:supp-pert02}
\begin{split}
\dot{q}_{k} = \frac{K_{k}}{n_k} \sum_{l=1}^{N} c_{kl} \int_0^{\infty} \du p_{k}(u) ~h^{\prime} \left[ \left( -\Omega(\tau_{kl}-\tau^{f}_{k}) - \beta_{kl} \right) \right] \left( q_{l}^{u+\tau_{kl}} - q_{k}^{u+\tau^{f}_{k}} \right).
\end{split}
\end{equation}
\noindent
Assuming exponential variations in the perturbations, \ie  $q_{k} = d_{k} e^{\lambda t}$, we get 
\begin{equation}\label{eq:supp-pert02}
\begin{split}
\lambda ~d_k e^{\lambda t}= \frac{K_{k}}{n_k} \sum_{l=1}^{N} c_{kl} \int_0^{\infty} \du p_{k}(u) 
~h^{\prime} \left[ \left( -\Omega(\tau_{kl}-\tau^{f}_{k}) - \beta_{kl} \right) \right] 
\left( d_{l} e^{-\lambda \tau_{kl}} - d_{k} e^{-\lambda \tau^f_{k}} \right) e^{\lambda t} e^{-\lambda u}.
\end{split}
\end{equation}
\noindent
For notation simplicity, denoting, $\tilde{c}_{kl} = c_{kl}/n_k$ and 
\begin{equation} \label{eq:supp-def_alpha}
\alpha_{kl} = K_{k} ~h^{\prime} \left[ \left( -\Omega(\tau_{kl}-\tau^{f}_{k}) - \beta_{kl} \right) \right], 
\end{equation}
\noindent 
above equation read, 
\begin{equation}\label{eq:supp-pert03}
\begin{split}
\lambda ~d_k = \sum_{l=1}^{N} \tilde{c}_{kl} \int_0^{\infty} \du p_{k}(u) ~\alpha_{kl}~  
\left( d_{l} e^{-\lambda \tau_{kl}} - d_{k} e^{-\lambda \tau^f_{k}} \right) e^{-\lambda u}.
\end{split}
\end{equation}
\noindent
Now using $\hat{p}_{k}(\lambda) = \int_0^{\infty} \du p_{k}(u) e^{-\lambda u}$, we have 
\begin{equation}\label{eq:supp-pert04}
\begin{split}
\lambda ~d_k =&~ \hat{p}_{k}(\lambda) \sum_{l=1}^{N} \tilde{c}_{kl} ~\alpha_{kl}~  
\left( d_{l} e^{-\lambda \tau_{kl}} - d_{k} e^{-\lambda \tau^{f}_{k}} \right);\\
\Rightarrow 
\frac{\lambda}{\hat{p}_{k}(\lambda) e^{-\lambda \tau^{f}_{k}}} ~d_k =&~ \sum_{l=1}^{N} \tilde{c}_{kl} ~\alpha_{kl}~  
\left( d_{l} e^{-\lambda (\tau_{kl} - \tau^{f}_{k})} - d_{k} \right);\\
\Rightarrow 
\left( \frac{\lambda}{\hat{p}_{k}(\lambda) e^{-\lambda \tau^{f}_{k}}} + 
\sum_{l=1}^{N} \tilde{c}_{kl} ~\alpha_{kl} \right) d_k =&~ 
\sum_{l=1}^{N} \left( \tilde{c}_{kl} ~\alpha_{kl}~e^{-\lambda (\tau_{kl} - \tau^{f}_{k})} \right) d_{l}. 
\end{split}
\end{equation}
\noindent
Therefore we find 
\begin{equation}\label{eq:supp-pert05}
\begin{split}
d_k =  
\sum_{l=1}^{N} \tilde{c}_{kl} ~\alpha_{kl} 
\left( \frac{e^{-\lambda (\tau_{kl} - \tau^{f}_{k})}}{\frac{\lambda}{\hat{p}_{k}(\lambda) e^{-\lambda \tau^{f}_{k}}} + f_k(\tilde{c},\alpha)} \right)
d_{l}, ~~~ k=1,2 \cdots, N,
\end{split}
\end{equation}
\noindent
where, 
\begin{equation} \label{eq:supp-def_fk}
f_k(\tilde{c},\alpha) = \sum_{l=1}^{N} \tilde{c}_{kl} ~\alpha_{kl}.
\end{equation}
\noindent 
In a matrix form we can write, 
\begin{equation} \label{eq:supp-mchar1}
\zeta \cdot \mathbf{d} = \mathbf{G} \cdot  \mathbf{d}, 
\end{equation}
\noindent
where, $\zeta = 1$, vector $\mathbf{d} = (d_1, d_2, d_3, \cdots, d_N)^{T}$ and the elements of the matrix $\mathbf{G}_{N \times N}$ are, 
\begin{equation} \label{eq:supp-jacobian2}
G_{ij} = \left( \frac{\tilde{c}_{ij} \alpha_{ij} e^{-\lambda (\tau_{ij} - \tau^{f}_{i})}}{\frac{\lambda}{\hat{p}_{i}(\lambda) e^{-\lambda \tau^{f}_{i}}} + f_i(\tilde{c},\alpha)} \right). 
\end{equation}
\noindent
From Eq.~\eqref{eq:supp-mchar1}, $\zeta$ ($= 1$) is the eigenvalue of the matrix $\mathbf{G}$ hence must satisfy the characteristic equation 
\begin{equation} \label{eq:supp-mchar2}
\det{ \left( \mathbf{G} - \zeta \cdot \mathbf{I} \right) } = 0,
\end{equation}
\noindent
Using above condition, for the network of $N$ delay-coupled PLLs, one can evaluate eigenvalues $\lambda$ of the perturbation dynamics and determine stability of the synchronized states. 


\section{Supplementary results with heterogeneous parameters}

\subsection{The effects of detuning}

As discussed in the paper, one of the effects of the detuned intrinsic frequencies is the appearance of the region in parameter space where synchronized solutions do not exist, see Fig.~\ref{fig:new-hf-fig14}. 
Here we examine how synchronized states in the system vanish as the detuning is increased. 
We find that the pairs of synchronized solutions go through a saddle node bifurcation as the magnitude of the detuning $\Delta\omega$ is increased from $\Delta\omega = 0$. 
This is shown in Supplementary Fig.~\ref{fig:new-hf-fig13}, where pair(s) of stable and unstable solution collide at the bifurcation point and disappear.  
Depending on the delay-values, this bifurcation can occur in one (see first and second column), or more pairs of solutions (third column), leading to the regimes in parameter space where synchronized solutions do not exist. 
\begin{figure}[h]
\includegraphics [scale=0.30, angle=0]{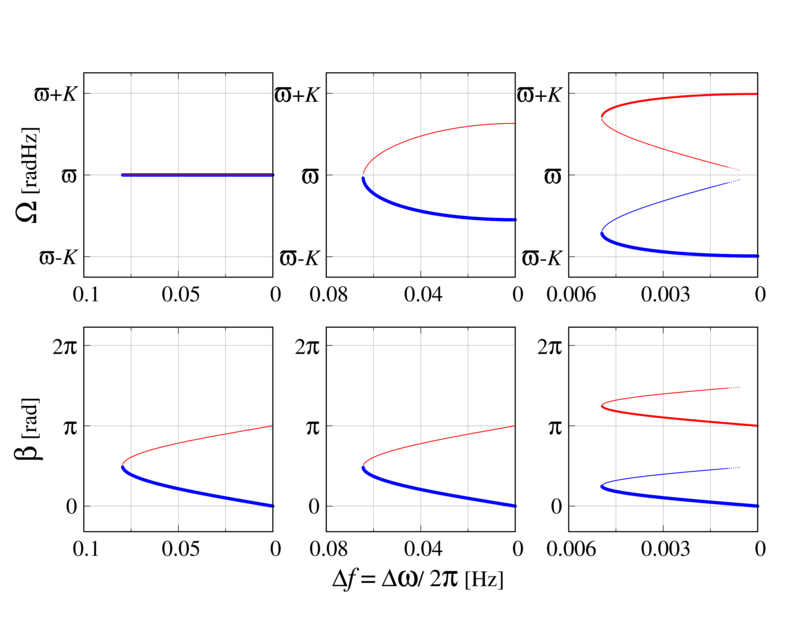}
 \caption{(Color online) Saddle-node bifurcation for one of the synchronized solutions (characterized by global frequency $\Omega$ and phase difference $\beta$) plotted as a function of intrinsic frequency detuning $\Delta f = \Delta \omega/2\pi \rm{~Hz}$. 
 The blue (dark gray) and red (light gray) curves denote near-inphase and near-antiphase synchronized solutions respectively. 
 Thick lines denote stable and thin unstable solutions. 
 Results for three different values of the transmission-delay $\tau = 0.25 \rm{~s}$ (left column), $0.35 \rm{~s}$ (middle column) and $0.5 \rm{~s}$ (right column) are shown. 
 Depending on the transmission-delay, a pair (or pairs) of synchronized solutions with opposite stability properties collide and disappear as the amount of detuning is increased.  
 The coupling strength $K = 0.25 \rm{~rad Hz}$ and the cut-off frequencies of the LFs $\omega_c = 0.25\times \bar{\omega} \rm{~rad Hz}$ are fixed. 
}
 \label{fig:new-hf-fig13}
\end{figure}

In Supplementary Fig.~\ref{fig:new-hf-fig12}, the behavior of the synchronized solutions is shown as a function of the transmission delay at a higher value of coupling strength as in Fig.~\ref{eq:n=2-stabcon2-h1} in section~\ref{subsec:detuning} in the paper. 
This figure shows that with increasing coupling strength, the transmission delay induced multistability~\cite{Schuster1989} and instabilities caused by the filtering process become more pronounced. 
With heterogeneous intrinsic frequencies $\Delta \omega \neq 0$ (right column), we can also observed the pair of additional solutions arising due to the symmetry breaking (see main text), which were absent when $\omega_1 = \omega_2$ (left column). 

\subsection{Heterogeneous transmission delays}

\begin{figure}
\includegraphics [scale=0.38, angle=0]{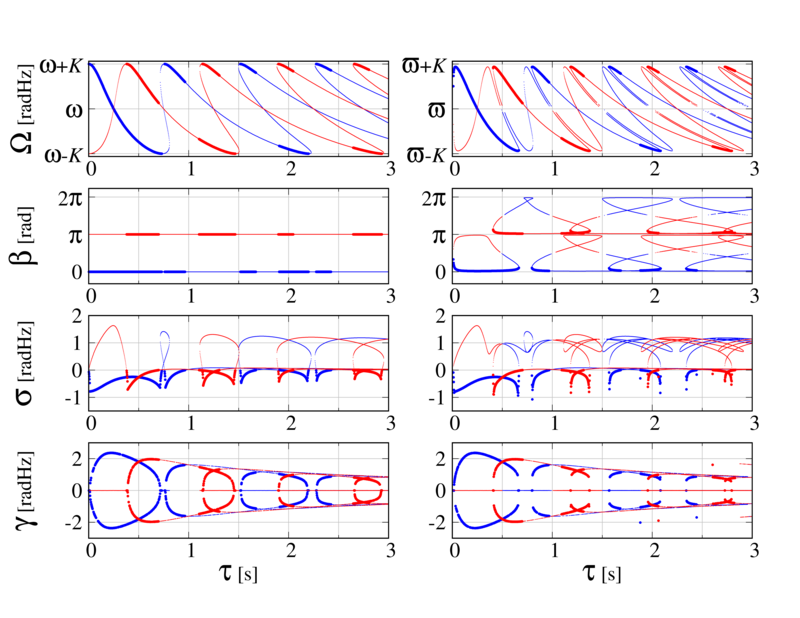}
 \caption{(Color online) Global frequency $\Omega$, phase configuration $\beta$ the perturbation response rate $\sigma$ and the corresponding modulation frequency $\gamma$ of the synchronized solutions plotted as a function of the transmission-delay $\tau$ for a system of two mutually delay-coupled PLLs at a coupling strength $K=2.0 \rm{~rad Hz}$. 
 All other parameters are the same as in Fig.~\ref{fig:new-hf-fig11} in the paper. 
 Additional synchronized solutions are visible for detuned intrinsic frequencies, see right column for $\Delta \omega = 0.04\times2\pi \rm{~rad Hz}$.  
}
\label{fig:new-hf-fig12}
\end{figure}
\begin{figure}[h]
\includegraphics [scale=0.32, angle=0]{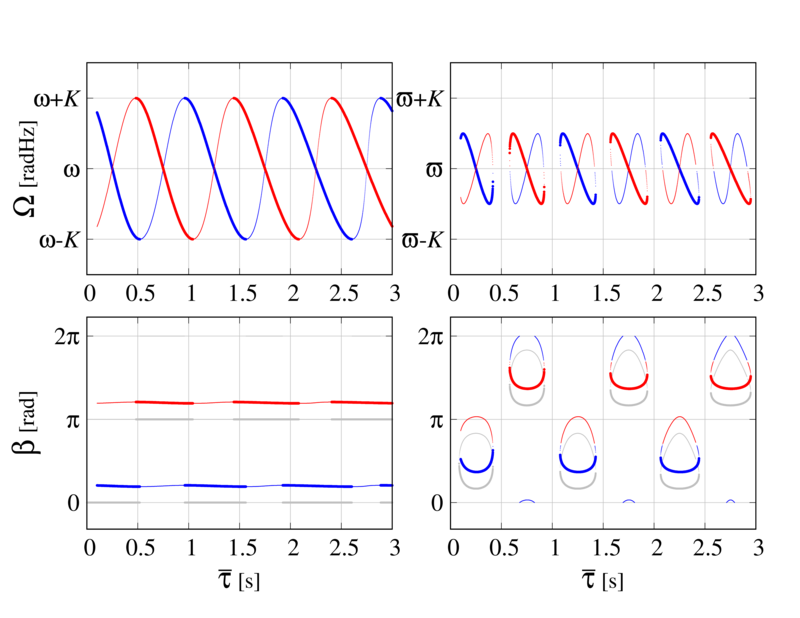}
 \caption{(Color online) Global frequency $\Omega$ and phase-configuration $\beta$ of synchronized states plotted as a function of the mean delay $\bar{\tau}$ with a constant delay-difference $\Delta\tau = -0.20 \rm{~s}$ for a system of two delay-coupled PLLs with identical $\omega_{1,2} = 1 \times 2\pi \rm{~rad Hz}$ (left column) and heterogeneous intrinsic frequencies $\omega_{1,2} = (1\mp0.02)\times 2\pi \rm{~rad Hz}$ (right column). 
 The blue (dark gray) and red (light gray) curves correspond to the inphase and antiphase ($\Delta\omega = 0\,\rm{radHz}$) or modified-inphase and modified-antiphase ($\Delta\omega = 0.04\times 2 \pi\,\rm{radHz}$) synchronized states.
 The thick and thin curves denote stable and unstable solutions respectively. 
 The phase configurations in the synchronized states without transmission-delay heterogeneity ($\Delta\tau = 0$) are plotted with gray curves for comparison. 
 The coupling strength is $K=0.25 \rm{~rad Hz}$ and the cut-off frequency is $\omega_c = 0.25\times \bar{\omega} \rm{~rad Hz}$. 
}
\label{fig:new-hf-dd-fig2}
\end{figure}

In the paper we discussed how the synchronized states are effected by the heterogeneity in the transmission delays. 
Specifically, for a fixed mean delay $\bar{\tau}$, we showed that the global frequencies of the synchronized state remain unaffected while the phase-difference changes linearly as a function of delay-heterogeneity $\Delta\tau$. 
Here we present additional results when the delay-heterogeneity $\Delta\tau$ is fixed and show the behavior of the system as a function of mean delay $\bar{\tau}$. 
In Fig.~\ref{fig:new-hf-dd-fig2} we plot the frequency and phase differences against the mean delay $\bar{\tau}$ for identical (left column $\Delta\omega = 0$) and detuned (right column $\Delta\omega = 0.04 \times 2 \pi$) intrinsic frequencies of the PLLs with a constant delay difference $\Delta \tau = -0.20\,\rm s$. 
We find that the frequencies of synchronized states of the system with homogeneous ($\tau_{12} = \tau_{21} = \bar{\tau}$) and heterogeneous delays $\tau_{1,2} = \bar{\tau} \pm \Delta\tau$ remain same for equal mean values $\bar{\tau}$. 
However, the phase-differences deviate from those in the case of homogeneous delay (gray curves for $\Delta\tau = 0$) by $-\Omega \Delta\tau/2$, see Eq.~\eqref{eq:n=2-sphase-sol-h2} in subsection~\ref{subsec:transmission-delay} of the paper.

\subsection{Heterogeneous coupling strengths}

\begin{figure}[h]
\includegraphics [scale=0.32, angle=0]{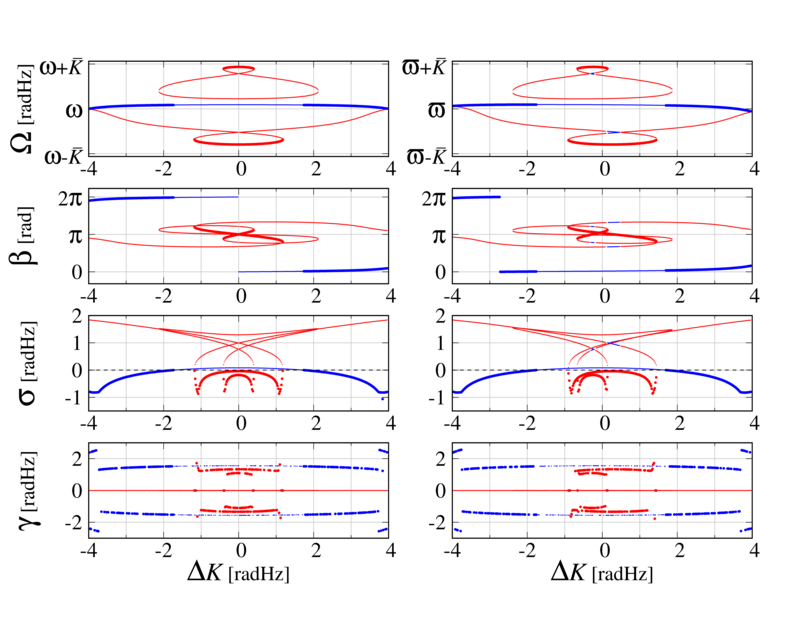}
 \caption{(Color online) 
Global frequency $\Omega$, phase-difference $\beta$, perturbation response rate $\sigma$ and the corresponding frequency $\gamma$ of the synchronized states as a function of differences in coupling strength $\Delta K = (K_2 - K_1)$ for two delay-coupled PLLs at fixed mean coupling $\bar{K}=2.0\,\rm{radHz}$, transmission-delay $\tau = 1.2~\rm{s}$ and cut-off frequency $\omega_c=0.25\times\bar{\omega}\,\rm{radHz}$. 
The blue (dark gray) and red (light gray) curves correspond to the inphase and antiphase ($\Delta\omega = 0\,\rm{radHz}$) or near-inphase and near-antiphase ($\Delta\omega = 0.04\times 2 \pi\,\rm{radHz}$) synchronized states.
The thick curves denote stable solutions (from Eq.~\eqref{eq:n=2-stabcon2-h1}, with $\sigma=\Re(\lambda_{max})<0$) and the thin curves unstable solutions.
The left column shows the results for $\omega_{1,2} = 1 \times 2\pi\,\rm{~rad Hz}$ and the right column the results for $\omega_{1,2} = (1 \mp 0.02) \times 2\pi\,\rm{radHz}$.
}
\label{fig:new-hc-fig4}
\end{figure}
In section~\ref{subsec:coupling}, we have discussed how the difference in coupling strengths $\Delta K$ affects the frequencies of synchronized states and phase difference in identical and detuned PLLs. 
We also observed how the coupling strength heterogeneity changes the perturbation decay rates of the synchronized solutions while the mean of the coupling values is kept constant (see Fig.~\ref{fig:new-hc-fig3}). 
In Supp. Fig.~\ref{fig:new-hc-fig4}, we present similar results for a larger value of the transmission delay, where many synchronized states coexist. 
For this case, we find for some of the synchronized solutions, the coupling heterogeneity may change the perturbation decay rates to the extant that it crosses zero value, \ie the stability of the solutions changes. 
This means that unstable synchronized solutions present for homogeneous coupling strength can be made stable as the system is tuned to heterogeneous coupling strength.

\end{widetext}




\begin{thebibliography}{39}
\bibitem[{eqv()}]{eqv-delay}
\bibinfo{note}{The values $\alpha_{12,21}$ in the characteristic equation
  depend on mean delays instead of their individual values. This can be shown
  by substituting $\tau_{12,21} = \bar{\tau} \mp \Delta\tau/2$ and using
  Eq.~\eqref{eq:n=2-all-hetero-sync-sol} with $K_1=K_2=K$. This gives 
  $\alpha_{12, 21} = -K \left[ \pm \sin(-\Omega \bar{\tau}) \sqrt{ 1- \left( 
  \frac{\Delta\omega}{2K \sin(\Omega\bar{\tau})}\right)^2 } \pm \frac{
  \Delta\omega \cos(-\Omega \bar{\tau})}{2K \sin(\Omega \bar{\tau})} \right]$.}
\end{thebibliography}
\end{document}